\documentclass[useAMS,usenatbib]{mn2e}

\usepackage{graphicx}

\def\aV{\mbox{$\rm A_V$}}

\def\jh{\mbox{$(J-H)$}}
\def\hk{\mbox{$(H-K_s)$}}
\def\jk{\mbox{$(J-K_s)$}}
\def\mMJ{\mbox{$(m-M)_J$}}
\def\mMo{\mbox{$(m-M)_O$}}
\def\ebv{\mbox{$E(B-V)$}}

\def\ejh{\mbox{$E(J-H)$}}

\def\rc{\mbox{$R_{\rm c}$}}

\def\rl{\mbox{$R_{\rm RDP}$}}
\def\rx{\mbox{$R_{\rm ext}$}}

\def\ms{\mbox{$M_\odot$}}
\def\ds{\mbox{$d_\odot$}}
\def\rs{\mbox{$R_\odot$}}
\def\dgc{\mbox{$R_{\rm GC}$}}

\def\dSC{\mbox{$\Delta R_{\rm SC}$}}

\def\jj{\mbox{$J$}}
\def\hh{\mbox{$H$}}
\def\ks{\mbox{$K_s$}}

\title[Properties of young and low-mass OCs]{The nature of the young and low-mass 
open clusters Pismis\,5, vdB\,80, NGC\,1931 and BDSB\,96}

\author[C. Bonatto and E. Bica]{C. Bonatto$^1$\thanks{E-mail: charles@if.ufrgs.br} and 
E. Bica$^1$\thanks{E-mail: bica@if.ufrgs.br}\\
$^1$ Departamento de Astronomia, Universidade Federal do Rio Grande do Sul\\ 
Av. Bento Gon\c{c}alves 9500, Porto Alegre 91501-970, RS, Brazil}

\begin{document}

\pagerange{\pageref{firstpage}--\pageref{lastpage}}

\maketitle

\label{firstpage}

\begin{abstract}
We investigate the nature of 4 young and low-mass open clusters (OCs) located in the 
$2^{nd}$ and $3^{rd}$ quadrants with near-IR 2MASS photometry (errors $<0.1$\,mag). 
After field decontamination, the colour-magnitude diagrams (CMDs) display similar 
morphologies: a poorly-populated main sequence (MS) and a dominant fraction of pre-MS 
(PMS) stars somewhat affected by differential reddening. Pismis\,5, vdB\,80 and BDSB\,96 
have MS ages within $5\pm4$\,Myr, while the MS of NGC\,1931 is $10\pm3$\,Myr old. However, 
non-instantaneous star formation is implied by the wider ($\sim20$\,Myr) PMS age spread. 
The cluster masses derived from MS$+$PMS stars are low, within $\sim60-180\,\ms$, with 
mass functions (MFs) significantly flatter than Salpeter's initial mass function (IMF). 
Distances from the Sun are within $1.0-2.4$\,kpc, and the visual absorptions are in the
range $\aV=1.0-2.0$. From the stellar radial density profiles (RDPs), we find that they 
are small ($\rc\la0.48$\,pc, $\rl\la5.8$\,pc), especially Pismis\,5 with $\rc\approx0.2$\,pc 
and $\rl\approx1.8$\,pc. Except for the irregular and cuspy inner regions of NGC\,1931 and 
Pismis\,5, the stellar RDPs follow a King-like profile. At $\sim10$\,Myr, central cusps - 
which in old clusters appear to be related to advanced dynamical evolution - are probably 
associated with a star-formation and/or molecular cloud fragmentation effect. Despite the 
flat MFs, vdB\,80 and BDSB\,96 appear to be typical young, low-mass OCs. NGC\,1931 and 
especially Pismis\,5, with  irregular RDPs, low cluster mass and flat MFs, do not appear 
to be in dynamical equilibrium. Both may be evolving into OB associations and/or doomed 
to dissolution in a few $10^7$\,yr.
\end{abstract}


\section{Introduction}
\label{Intro}

The first few ten Myrs represent the most critical phase in the life of a star 
cluster, to the point that only about 5\% (e.g. \citealt{LL2003}) of the embedded 
clusters are able to dynamically evolve into bound open clusters (OCs). The rapid 
gas removal by supernovae and massive star winds associated with this period can 
produce important changes on the primordial gravitational potential. Obviously, 
this effect depends essentially on the star-formation efficiency, the actual mass 
of primordial gas converted into stars and the mass of the more massive stars.

Because of the rather rapidly-reduced potential, a significant fraction of the
stars - the low mass ones in particular - moving faster than the scaled-down escape 
velocity may be driven into the field. Over a time-scale of $10-40$\,Myr, this 
process in turn, can dissolve most of the very young star clusters (e.g. 
\citealt{GoBa06}). 

Observationally, low-mass star clusters younger than about 10\,Myr present an
under-populated, developing main sequence (MS) and a more conspicuous population of 
pre-MS (PMS) stars. Typical examples are NGC\,6611 (\citealt{N6611}), 
NGC\,4755 (\citealt{N4755}), NGC\,2239 and NGC\,2244 (\citealt{N2244}) and Bochum\,1 
(\citealt{Bochum1}). On a large scale, the important changes to the potential that 
affect the early cluster spatial structure should also be reflected on the stellar 
radial density profile (RDP). Bochum\,1 and NGC\,2244, for instance, can be taken 
as representative of this scenario, in which an irregular RDP cannot be represented 
by a cluster-like (i.e. an approximately isothermal sphere) profile. Such conspicuous
RDP irregularities suggest significant profile erosion or dispersion of stars 
from a primordial cluster (e.g. \citealt{Bochum1}; \citealt{N2244}). 

In some cases, early star cluster dissolution may lead to the formation of low-mass 
OB associations, the subsequent dispersion of which may be an important source of
field stars (e.g. \citealt{Massey95}). Bochum\,1, with an irregular and clumpy RDP, 
can be an example of such an evolving structure. In this context, NGC\,2244 appears 
to be another example of a young OC in the process of dissolving in a few $10^7$\,yr. 
Perhaps the difference between objects like Bochum\,1 and NGC\,2244 and {\em normal} 
young OCs (i.e. with the combined population of MS and PMS stars distributed 
according to a cluster RDP as in 
NGC\,6611 and NGC\,4755) is related to the interplay between environment conditions, 
star-formation efficiency and stellar mass. Interestingly, the mass stored in the
MS$+$PMS members of Bochum\,1 and NGC\,2244 is a factor 2-3 lower than those of NGC\,6611 
and NGC\,4755.

In this paper we investigate the nature of the poorly-studied, young and low-mass 
OCs Pismis\,5, vdB\,80, NGC\,1931 and BDSB\,96, by means of their photometric and 
structural properties. Their location in the Galaxy, in the $2^{nd}$ and $3^{rd}$ 
quadrants (Table~\ref{tab1}), minimises field-star contamination (e.g. \citealt{ProbFSR}; 
\citealt{AntiC}), which is essential when PMS stars are expected to dominate in 
Colour-Magnitude Diagrams (CMDs). Our main goal is to determine whether such young 
and low-mass systems can be characterised as typical OCs or if they are on their 
way to dissolution. In addition, we will derive their fundamental and structural 
parameters, most of these for the first time.

\begin{figure}
\begin{minipage}[b]{0.50\linewidth}
\includegraphics[width=\textwidth]{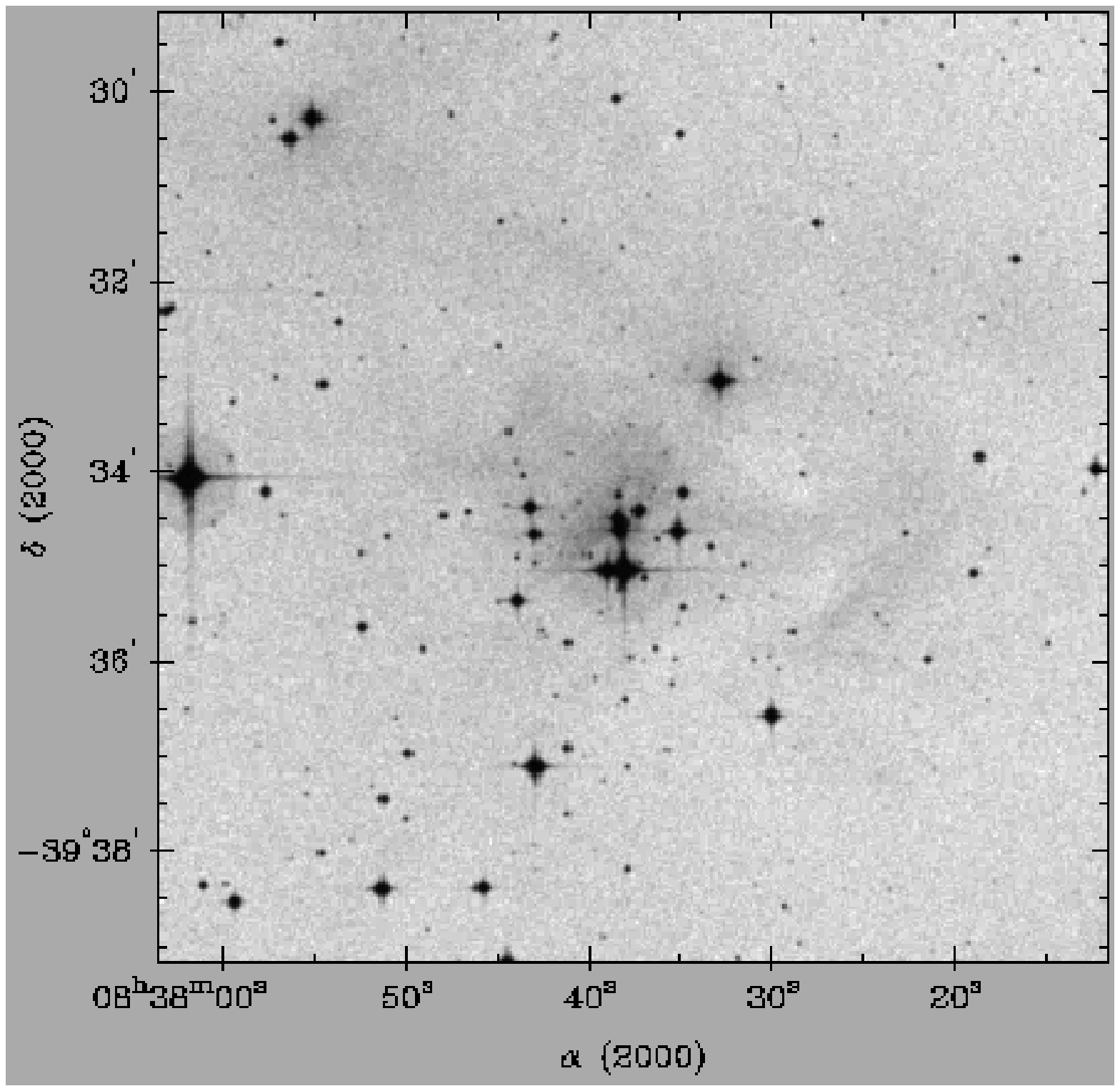}
\end{minipage}\hfill
\begin{minipage}[b]{0.50\linewidth}
\includegraphics[width=\textwidth]{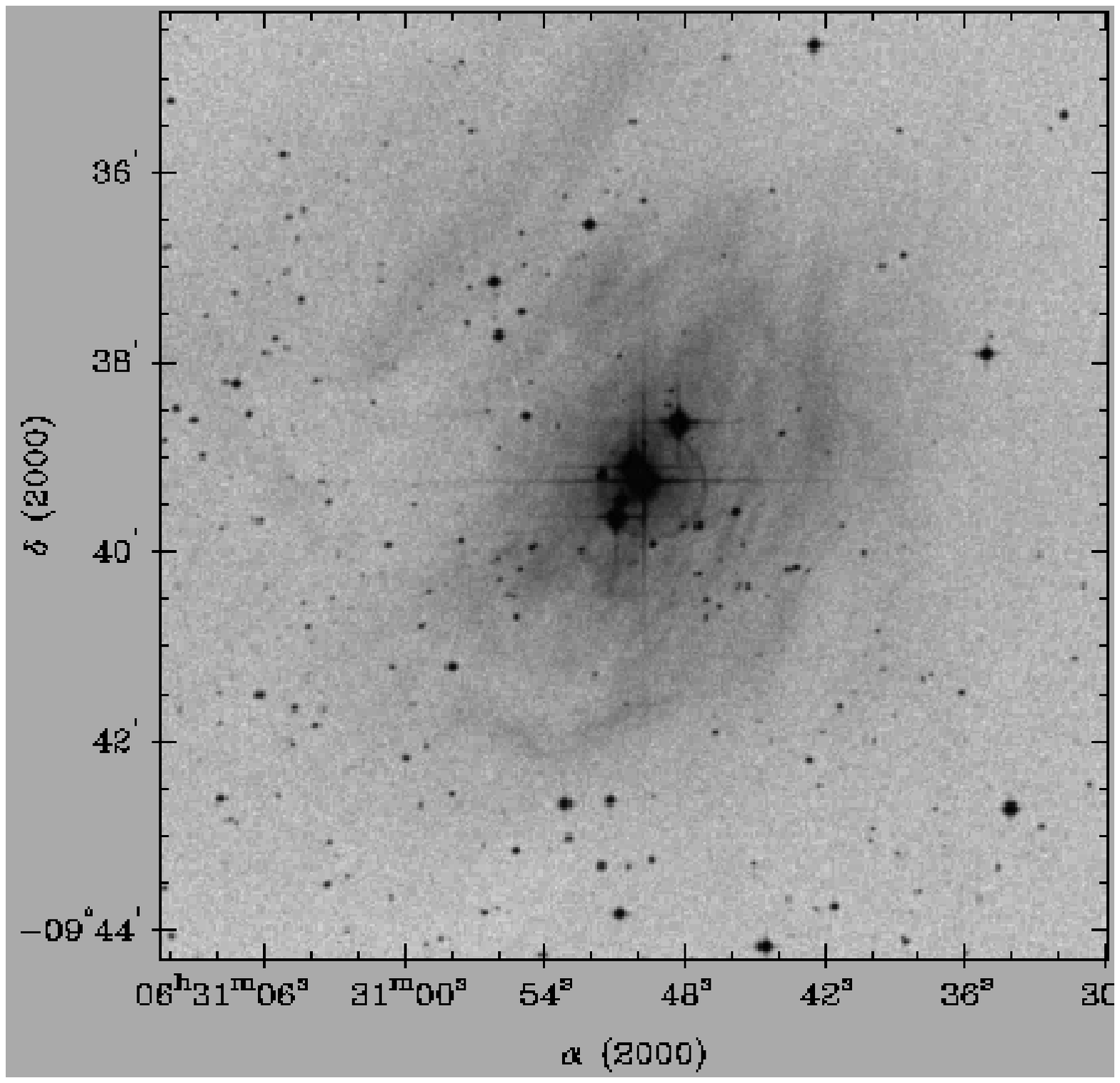}
\end{minipage}\hfill
\begin{minipage}[b]{0.50\linewidth}
\includegraphics[width=\textwidth]{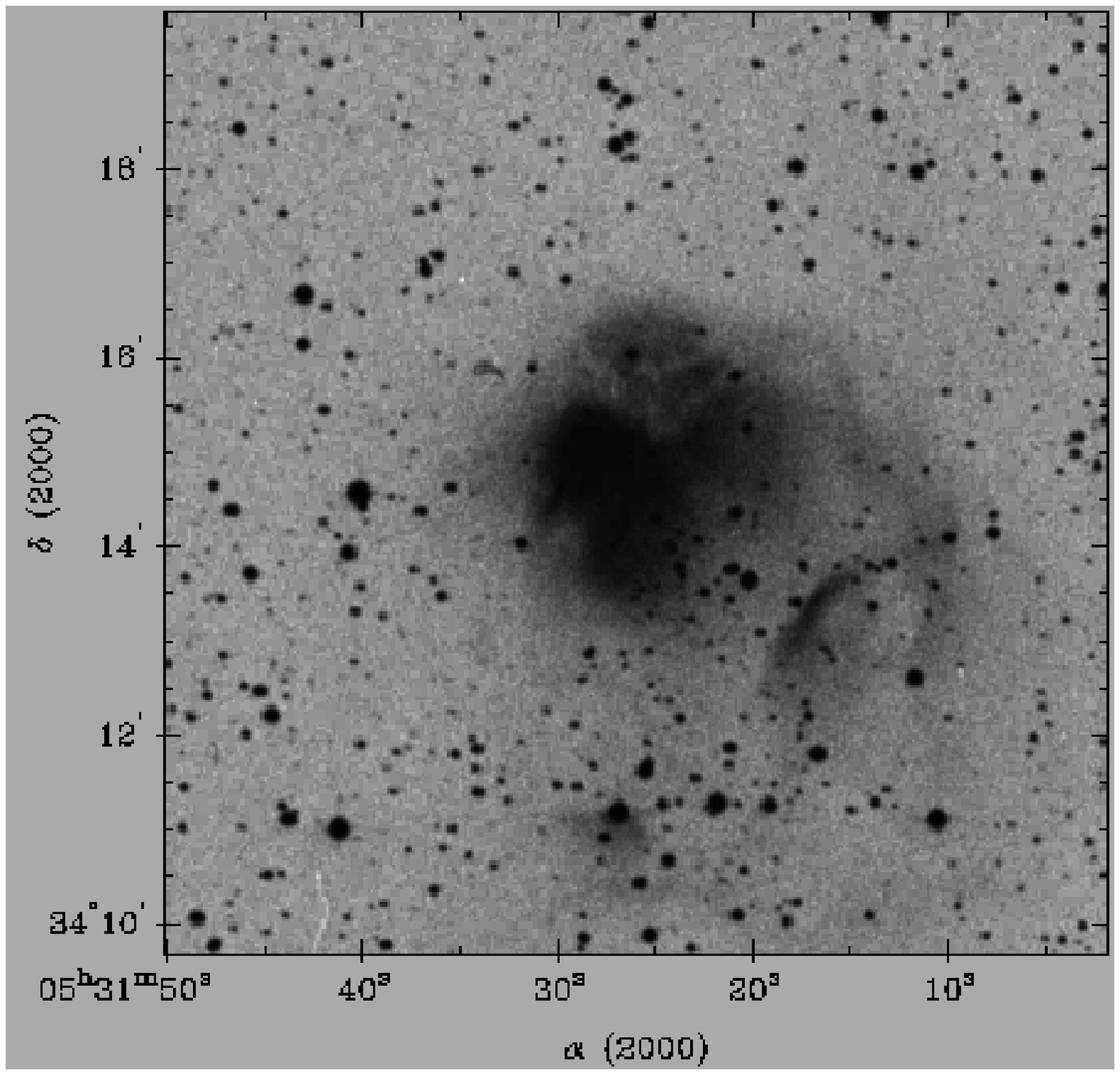}
\end{minipage}\hfill
\begin{minipage}[b]{0.50\linewidth}
\includegraphics[width=\textwidth]{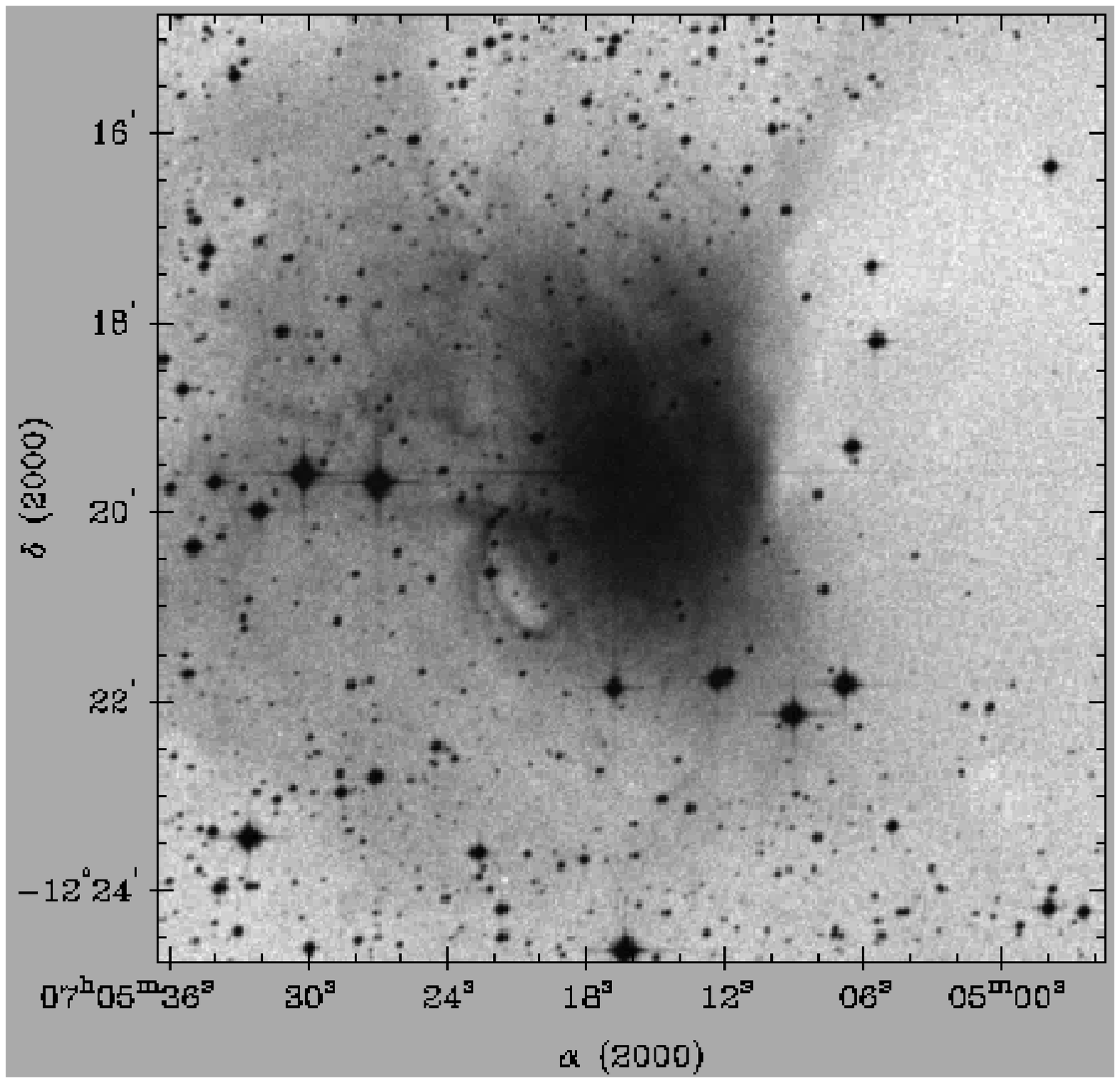}
\end{minipage}\hfill
\caption[]{The optical environments are shown in $10\arcmin\times10\arcmin$ DSS 
B images of Pismis\,5 (top-left), vdB\,80 (right), NGC\,1931 (bottom-left) and 
BDSB\,96 (right). Gas emission, dust reflection and/or absorption are present 
in all fields in varying proportions. Orientation: North to the top and East to 
the left.}
\label{fig1}
\end{figure}

\begin{figure}
\begin{minipage}[b]{0.50\linewidth}
\includegraphics[width=\textwidth]{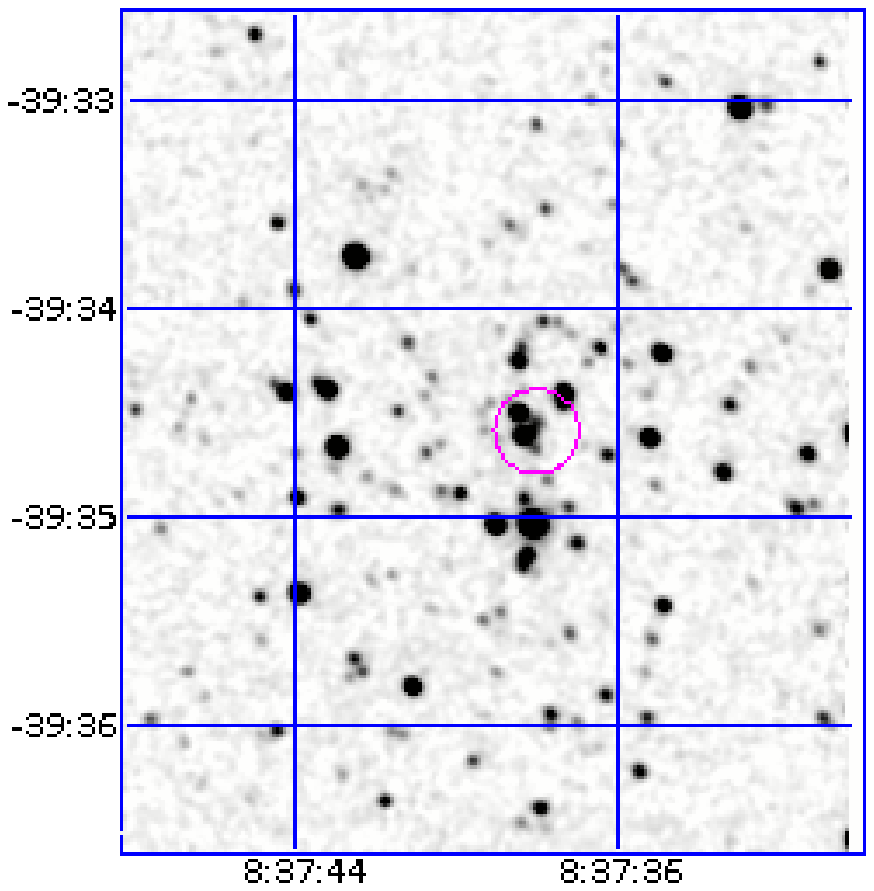}
\end{minipage}\hfill
\begin{minipage}[b]{0.50\linewidth}
\includegraphics[width=\textwidth]{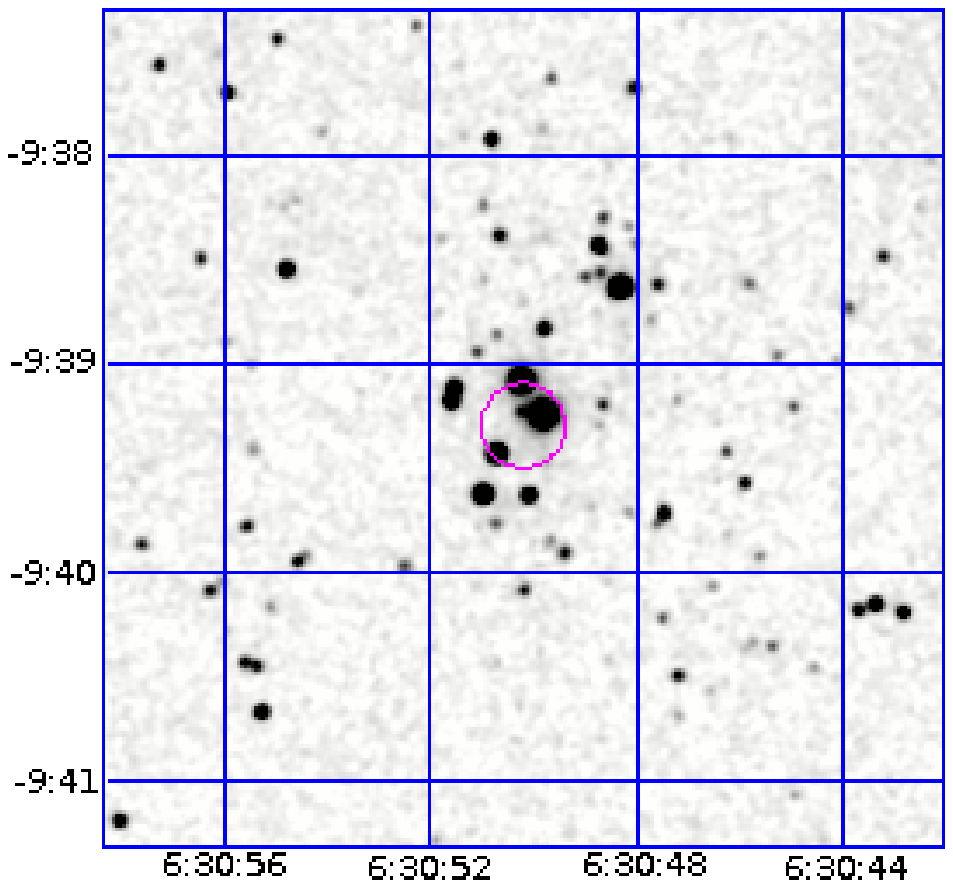}
\end{minipage}\hfill
\begin{minipage}[b]{0.50\linewidth}
\includegraphics[width=\textwidth]{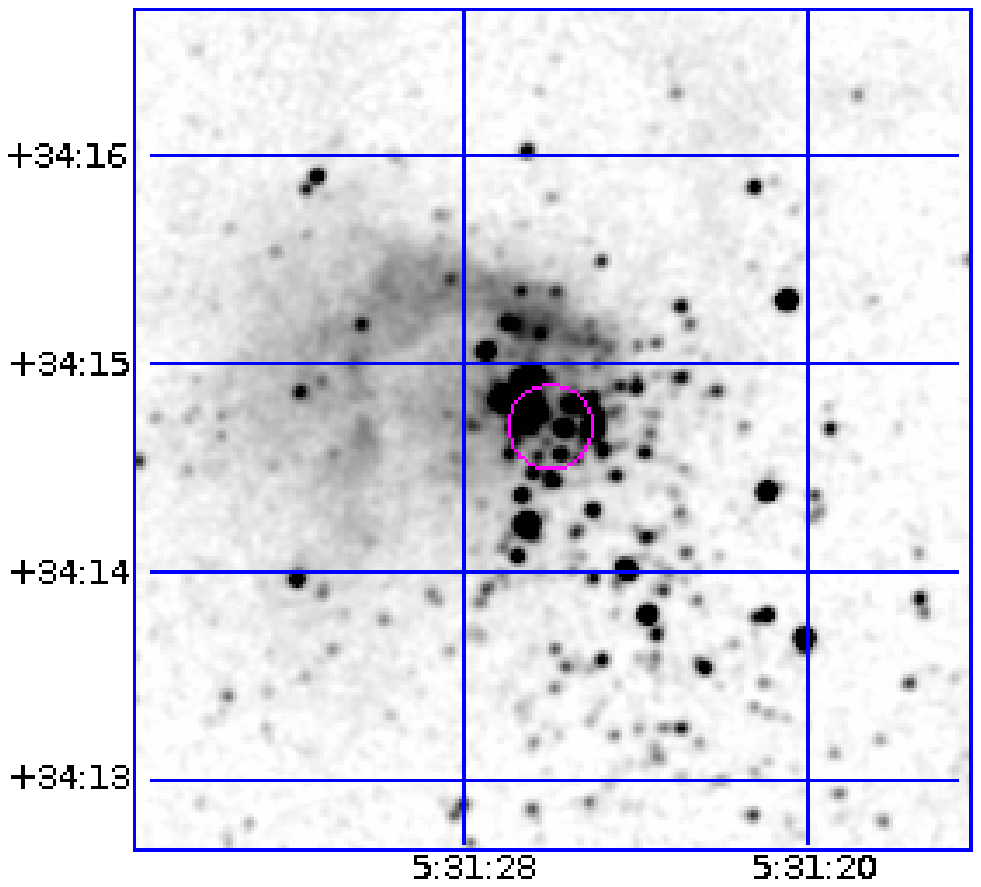}
\end{minipage}\hfill
\begin{minipage}[b]{0.50\linewidth}
\includegraphics[width=\textwidth]{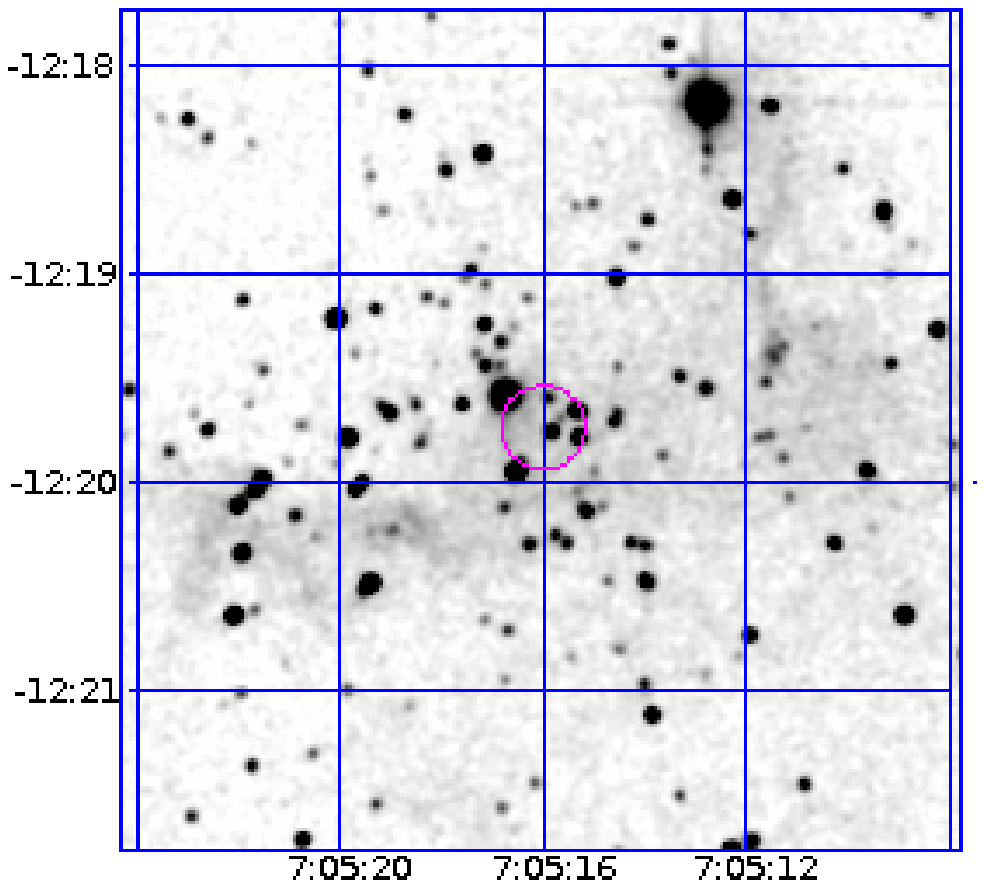}
\end{minipage}\hfill
\caption[]{$4\arcmin\times4\arcmin$ 2MASS \ks\ images showing the embedded clusters 
Pismis\,5 (top-left), vdB\,80 (right), NGC\,1931 (bottom-left) and BDSB\,96 (right).  
Right ascension and declination are given in the abscissa and ordinate, respectively.
Orientation: North to the top and East to the left. }
\label{fig2m}
\end{figure}

\begin{table*}
\caption[]{Fundamental parameters of the target OCs}
\label{tab1}
\tiny
\renewcommand{\tabcolsep}{0.95mm}
\renewcommand{\arraystretch}{1.25}
\begin{tabular}{cccccccccccccccccc}
\hline\hline
&\multicolumn{7}{c}{Literature}&&\multicolumn{9}{c}{This work}\\
\cline{2-8}\cline{10-18}
Cluster&$\alpha(2000)$&$\delta(2000)$&Age&\ebv&\ds&D&Source&&$\alpha(2000)$&$\delta(2000)$&
        $\ell$&$b$&Age&\ebv&\ds&\dSC&\rx\\
 & (hms)&($\degr\,\arcmin\,\arcsec$)&(Myr)&(mag)&(kpc)&(\arcmin)&&&(hms)&($\degr\,\arcmin\,\arcsec$)
 &(\degr)&(\degr)&(Myr)&(mag)&(kpc)&(kpc)&(\arcmin)\\
(1)&(2)&(3)&(4)&(5)&(6)&(7)&(8)&&(9)&(10)&(11)&(12)&(13)&(14)&(15)&(16)&(17)\\
\hline
Pismis\,5&08:37:38&$-$39:35:00&15&0.42&0.87&5.0&1&&08:37:37.5&$-$39:34:10.5&259.33&$+$0.93
         &$5\pm4$&$0.42\pm0.03$&$1.0\pm0.1$&$+0.3\pm0.1$&40 \\
vdB\,80&06:30:48&$-$09:40:00&4.5&0.38&---&5.0&1&&06:30:50.2&$-$09:39:18.0&219.26&$-$8.93
       &$5\pm2$&$0.61\pm0.10$&$2.1\pm0.3$&$+1.7\pm0.3$&40 \\
NGC\,1931&05:31:25&$+$34:14:32&10&0.74&3.1&4.0&1&&05:31:24.7&$+$34:14:41.2&173.90&$+$0.28
         &$10\pm3$&$0.62\pm0.10$&$2.4\pm0.3$&$+2.4\pm0.4$&60 \\
BDSB\,96&07:05:18&$-$12:19:44&---&---&1.1&2.7&2&&07:05:17.3&$-$12:19:9.5&225.46&$-$2.57
         &$5\pm3$&$0.37\pm0.10$&$1.4\pm0.2$&$+1.0\pm0.2$&60 \\
\hline
\end{tabular}
\begin{list}{Table Notes.}
\item Sources: (1) - WEBDA\footnote{\em obswww.univie.ac.at/webda}; (2) - \citet{BDS2003}. 
Col.~7: Optical diameter; Col.~15: distance from the Sun; col.~16: distance from the Solar 
circle; col.~17: extraction radius.
\end{list}
\end{table*}

This paper is organised as follows. In Sect.~\ref{RecAdd} we recall literature data on 
the target objects, discuss the 2MASS photometry and build the field-star decontaminated 
CMDs. In Sect.~\ref{DFP} we derive fundamental cluster parameters. In Sect.~\ref{struc} 
we derive structural parameters. In Sect.~\ref{MF} we estimate cluster mass. In 
Sect.~\ref{Discus} we compare  structural parameters and dynamical state with those of a 
sample of nearby OCs. Concluding remarks are given in Sect.~\ref{Conclu}.

\section{The OCs and the 2MASS photometry}
\label{RecAdd}

The optical environments of the objects are shown in $10\arcmin\times10\arcmin$ 
DSS\footnote{Extracted from 
the Canadian Astronomy Data Centre (CADC) - \em http://cadcwww.dao.nrc.ca/} B 
images (Fig.~\ref{fig1}), in which nebular gas and/or dust emission is present
in varying proportions. The embedded clusters show up in the $4\arcmin\times4\arcmin$ 
2MASS \ks\ images (Fig.~\ref{fig2m}). 

Pismis\,5 (also known as ESO\,313-SC7) is located in Vela and its field contains a 
few relatively bright stars mixed with gas emission and dust (Fig.~\ref{fig1}, 
top-left panel). It was discovered by P. Pismis (\citealt{Pismis59}) on Schmidt 
plates from the Tonantzintla Observatory (Mexico). According to 
SIMBAD\footnote{http://simbad.u-starsbg.fr/simbad}, the 2 brightest stars in this
field are $CD-39~4582$ ($\alpha(2000) = 08^h37^m38.0^s$, $\delta(2000) = 
-39\degr35\arcmin02\arcsec$, $V=9.9$, $\jj=9.4$, $\ks=9.1$) and $CD-39~4599$ 
($\alpha(2000) = 08^h38^m1.8^s$, $\delta(2000) = -39\degr34\arcmin6.1\arcsec$, 
$V=10.0$, $\jj=10.0$, $\ks=9.9$). Both appear to be members of Pismis\,5 
(Fig.~\ref{fig4}). As far as we are aware, the only work on Pismis\,5 is the 
UBV$+H_\beta$ photometric survey by \citet{VM73}, in which it was not considered 
as a star cluster. 

Cradled inside the reflection nebula vdB-RN\,80 (\citealt{vdB66}) in Monoceros,
the field of the star cluster vdB\,80 (Fig.~\ref{fig1}, top-right panel) is similar 
to that of Pismis\,5. The brightest star $NSV~2998$ 
(SIMBAD: $\alpha(2000) = 06^h30^m49.8^s$, $\delta(2000) = -09\degr39\arcmin14.8\arcsec$, 
$V=8.8$, $\jj=8.1$, $\ks=8.1$) appears to be the only MS member of vdB\,80 
(Fig.~\ref{fig4}). In the only reference to vdB\,80 as a star cluster, \citet{Ahumada01} 
estimated the age $4.5\pm1.5$\,Myr by means of integrated spectra.

The star cluster NGC\,1931 (Stock\,9, Collinder\,68) lies within 
the bright nebula NGC\,1931 (Sh\,2-237) in Auriga. \citet{MFJ79} determined the distance 
from the Sun $\ds=1.8$\,kpc, while \citet{PM86} found $\ds=2.2$\,kpc. \citet{Bhatt94} 
found $\ds=2.2$\,kpc, $\ebv=0.55$ and the age $\sim10$\,Myr. \citet{Lata02} derived 
$\ebv=0.58$ and $\sim13$\,Myr. Finally, \citet{CCS04} found $\ds=3.1$\,kpc, $\sim10$\,Myr 
and the cluster radius $R=4\farcm13$. The latter work was based on 2MASS photometry, as 
in the present study with our tools, since the cluster is embedded in nebular emission.

Discovered in a survey of infrared embedded star clusters and stellar groups carried with 
2MASS by \citet{BDS2003}, BDSB\,96 (in the nebula Cederblad\,90, also known as Gum\,3, 
Sh\,2-297, RCW\,1a, vdB-RN\,94 and Ber\,134) in Canis Major, was classified as an infrared 
OC located at about 1.1\,kpc from the Sun. Its field contains the single bright star 
$HD~53623$ (SIMBAD: $\alpha(2000) = 07^h05^m16.7^s$, $\delta(2000) = -12\degr19\arcmin34.5\arcsec$, 
$V=8.0$, $\jj=8.0$, $\ks=8.0$), which is in the MS of BDSB\,96 (Fig.~\ref{fig5}). 
Nebular gas and/or dust emission is more conspicuous in its field (Fig.~\ref{fig1}, 
bottom-right panel) than in Pismis\,5, vdB\,80 and NGC\,1931.

Table~\ref{tab1} provides parameters found in the literature and derived here.
Since accurate spectral types of the bright stars listed above are not available,
spectroscopic distances cannot be computed. The central coordinates were recomputed 
to match the absolute maximum present in the stellar surface densities (Sect.~\ref{DecOut}).

\subsection{2MASS photometry}
\label{2mass}

The present OCs still retain part of the primordial gas and dust (Fig.~\ref{fig1}), 
which makes the near-IR the optimal spectral range to probe them. For instance, the 
number of detected stars at a given radius in NGC\,2244 in the optical is significantly 
lower than in the near-IR, dropping to about 5\% for the full field (\citealt{N2244}). 
For this purpose we work with the 2MASS\footnote{The Two Micron All Sky Survey, All 
Sky data release (\citealt{2mass1997}) - {\em http://www.ipac.caltech.edu/2mass/releases/allsky/}} 
near-IR \jj, \hh, and \ks\ photometry, which allows the spatial and photometric 
uniformity useful for wide extractions that, in turn, provide high star-count 
statistics. Within this perspective, we have been developing quantitative tools to 
statistically disentangle cluster evolutionary sequences from field stars in CMDs. 
Decontaminated CMDs, in turn, have been used to investigate the nature of cluster 
candidates and derive their astrophysical parameters (e.g. \citealt{ProbFSR}). In short, 
we apply {\em (i)} field-star decontamination to uncover the intrinsic CMD morphology, 
essential for a proper derivation of reddening, age, 
and distance from the Sun, and {\em (ii)} colour-magnitude filters, which are 
essential for intrinsic stellar RDPs, as well as luminosity and mass functions (MFs). 
In particular, the use of field-star decontamination in the construction of CMDs has 
proved to constrain age and distance more than working with raw (observed) photometry, 
especially for low-latitude OCs (\citealt{discProp}).

2MASS can reach adequate CMD depths for nearby OCs. For instance, our group has studied 
the young OCs NGC\,6611, NGC\,4755, NGC\,2239 and NGC\,2244. Abundant PMS stars were 
seen in the $\approx1$\,Myr old NGC\,6611 (which is essentially embedded) and the 
$1-6$\,Myr old NGC\,2244, and a few remaining ones in the $\approx14$\,Myr old NGC\,4755. 
As nearby older OCs we cite NGC\,2477 (\citealt{DetAnalOCs}) and M\,67 (\citealt{M67}). 

Photometry was extracted in a wide circular field of radius \rx\ (Table~\ref{tab1}) with
VizieR\footnote{\em http://vizier.u-strasbg.fr/viz-bin/VizieR?-source=II/246}. 
The extraction radii are large enough to allow determination of the background level 
(Sect.~\ref{struc}) and to statistically characterise the colour and magnitude distribution
of the field stars (Sect.~\ref{Decont_CMDs}). As photometric quality constraint, only stars 
with \jj, \hh, and \ks\ errors lower than 0.1\,mag were considered\footnote{In all cases,
the bright stars have photometric errors within the adopted range. Besides, any bias against 
low-mass stars (larger errors) introduced by this criterion (applied equally for cluster and
field stars) is probably minimised by the field decontamination (Sect.~\ref{Decont_CMDs}).}.  
Reddening corrections are based on the relations $A_J/A_V=0.276$, $A_H/A_V=0.176$, $A_{K_S}/A_V=0.118$, 
and $A_J=2.76\times\ejh$ given by \citet{DSB2002}, with $R_V=3.1$, considering the extinction 
curve of \citet{Cardelli89}.

\subsection{Field decontamination}
\label{Decont_CMDs}

Field-star decontamination is a very important step in the identification and characterisation 
of star clusters, especially the clusters near the Galactic equator. Different approaches
are described in \citet{N2244}. In this paper we employ the decontamination algorithm detailed in 
\citet{BB07}, \citet{ProbFSR} and \citet{F1603}, and briefly described below. 

\begin{figure}
\resizebox{\hsize}{!}{\includegraphics{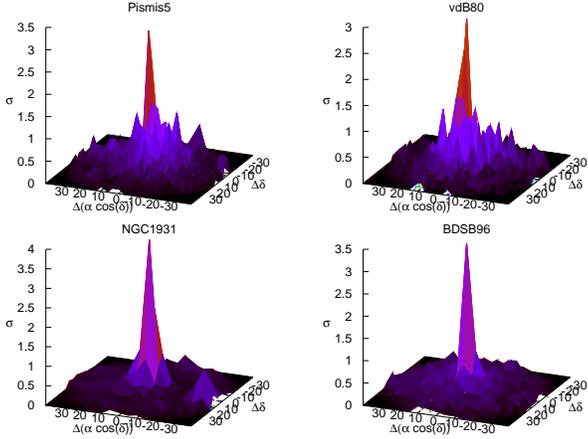}}
\caption[]{Stellar surface-density $\sigma(\rm stars\ arcmin^{-2})$ computed with
field-decontaminated photometry to enhance the cluster/background contrast. 
$\Delta(\alpha~\cos(\delta))$ and $\Delta\delta$ in arcmin.}
\label{fig2}
\end{figure}

\begin{figure}
\resizebox{\hsize}{!}{\includegraphics{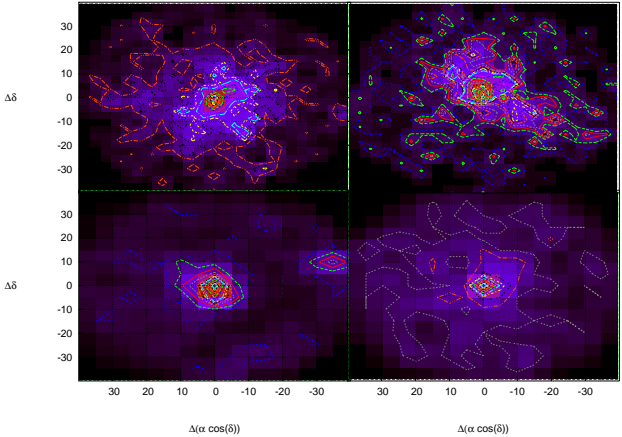}}
\caption[]{Similar to Fig.~\ref{fig2} for the isopleth curves, for a better visualisation 
of cluster extent and geometry.}
\label{fig3}
\end{figure}

The algorithm {\em (i)} divides the whole range of CMD magnitude and colours into a 3D 
grid of cells with axes along the \jj\ magnitude and the \jh\ and \jk\ colours, {\em (ii)} 
estimates the number density of field stars in each cell based on the number of comparison 
field stars with similar magnitude and colours as those in the cell, and {\em (iii)} subtracts
the estimated number of field stars from each cell. Typical cell dimensions are $\Delta\jj=1.0$ 
and $\Delta\jh=\Delta\jk=0.2$; the comparison fields are located within $R=30\arcmin$ and
the extraction radius. With this setup, the subtraction efficiency, i.e. the difference between
the background contamination (which may be fractional) and the number of subtracted stars in 
each cell (e.g., \citealt{AntiC}), summed over all cells, is higher than 90\% in all cases.

\subsection{Decontaminated surface density maps}
\label{DecOut}

Figure~\ref{fig2} shows the spatial distribution of the stellar surface-density ($\sigma$, in 
units of $\rm stars\,arcmin^{-2}$) around the 4 objects, measured with 2MASS photometry. We use 
field-star decontaminated photometry (Sect.~\ref{Decont_CMDs}) to maximise the cluster/background 
contrast. The surface density is computed in a rectangular mesh with cells 
$2.5\arcmin\times2.5\arcmin$ wide, reaching total offsets of $|\Delta(\alpha~\cos(\delta))|=|\Delta\delta|\approx40\arcmin$ with respect to the cluster centre 
(Table~\ref{tab1}). Despite the gas and dust associated with the present low-mass clusters
(Fig.~\ref{fig1}), the central excesses are conspicuously detected in the decontaminated 
surface-density distributions. Besides, the residual surface-density around the centre has 
been reduced to a minimum level.

By design, our decontamination depends essentially on the colour-magnitude distribution of 
stars located in different spatial regions. The fact that the decontaminated surface-density 
presents a conspicuous excess only at the assumed cluster position implies significant differences 
among this region and the comparison field, both in terms of colour-magnitude and star counts 
within the corresponding colour-magnitude bins. This meets cluster expectations, which can be 
characterised by a single-stellar population, projected against a Galactic stellar field. 

The respective isopleth surface maps are shown in Fig.~\ref{fig3}, in which cluster size 
and geometry can be observed. In all cases, the central region ($R<5\arcmin$) appears 
essentially circular, with elongated external regions, especially NGC\,1931.

\section{Fundamental parameters}
\label{DFP}

\begin{figure}
\resizebox{\hsize}{!}{\includegraphics{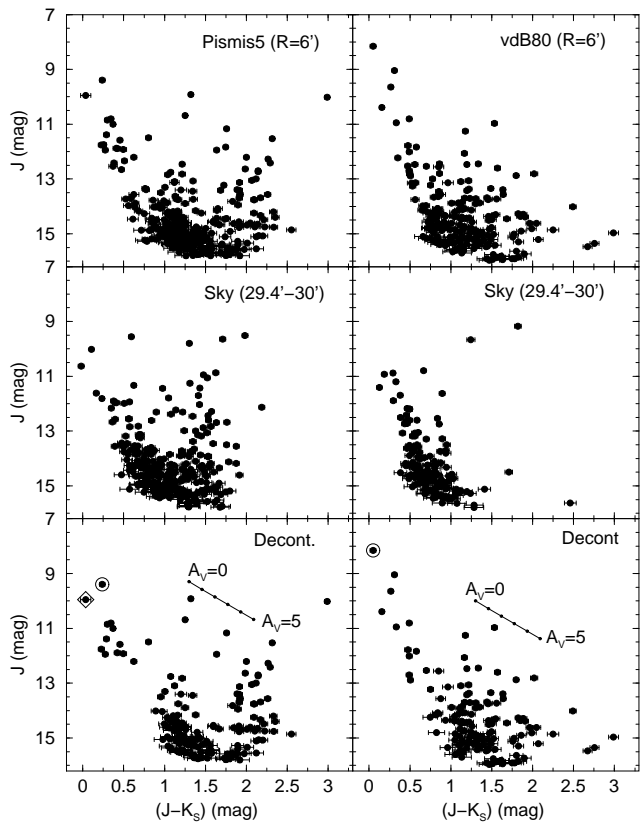}}
\caption[]{2MASS CMDs of Pismis\,5 (left) and vdB\,80 (right). Top panels: observed photometry. 
Middle: equal-area comparison field CMDs. Bottom: decontaminated CMDs. Bright stars in common 
with SIMBAD are: $CD-39~4582$ (large circle) and $CD-39~4599$ (diamond) in Pismis\,5, and  
$NSV~2998$ (large circle) in vdB\,80. Reddening vectors for $\aV=0-5$ are shown in the bottom 
panels.}
\label{fig4}
\end{figure}

\begin{figure}
\resizebox{\hsize}{!}{\includegraphics{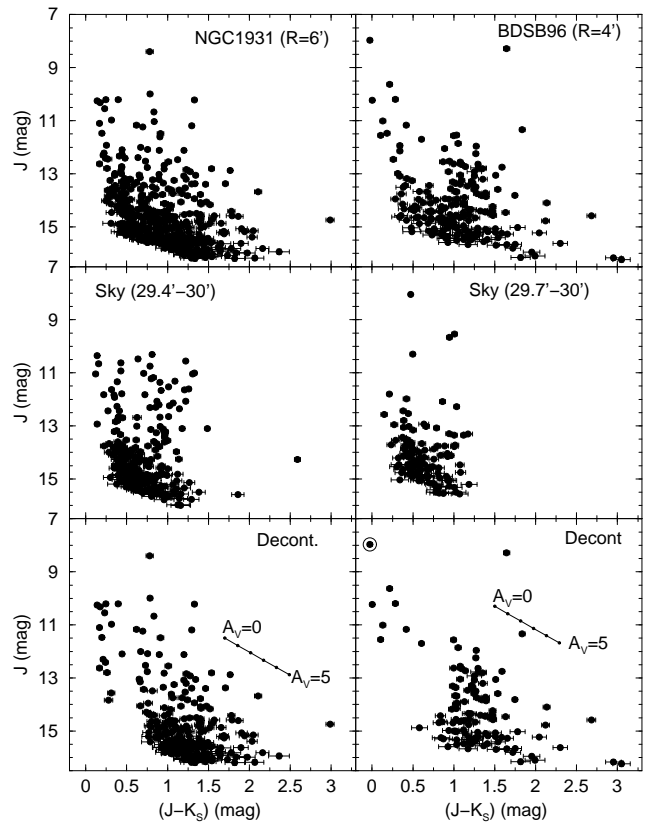}}
\caption[]{Similar to Fig.~\ref{fig4} for NGC\,1931 (left) and BDSB\,96 (right). The 
bright star in BDSB\,96 is $HD~53623$.}
\label{fig5}
\end{figure}

$\jj\times\jk$ CMDs built with the raw photometry of the sample objects are shown 
in the top panels of Figs.~\ref{fig4} and \ref{fig5}. In all cases, the sampled region 
contains most of the cluster stars (Fig.~\ref{fig9}). When qualitatively compared with 
the CMDs extracted from the equal-area comparison field\footnote{The equal-area field 
extractions serve only for qualitative comparisons, since the decontamination uses a 
large surrounding area (Sect.~\ref{Decont_CMDs}).} (middle panels), features typical 
of very young OCs emerge: a relatively vertical and poorly-populated MS together with a
large number of faint and red PMS stars.

The decontaminated CMDs are shown in the bottom panels of Figs.~\ref{fig4} and 
\ref{fig5}. As expected, essentially all contamination is removed, leaving stellar 
sequences typical of mildly reddened, young and low-mass OCs, with a developing MS 
and a significant population of PMS stars. 

\begin{figure}
\resizebox{\hsize}{!}{\includegraphics{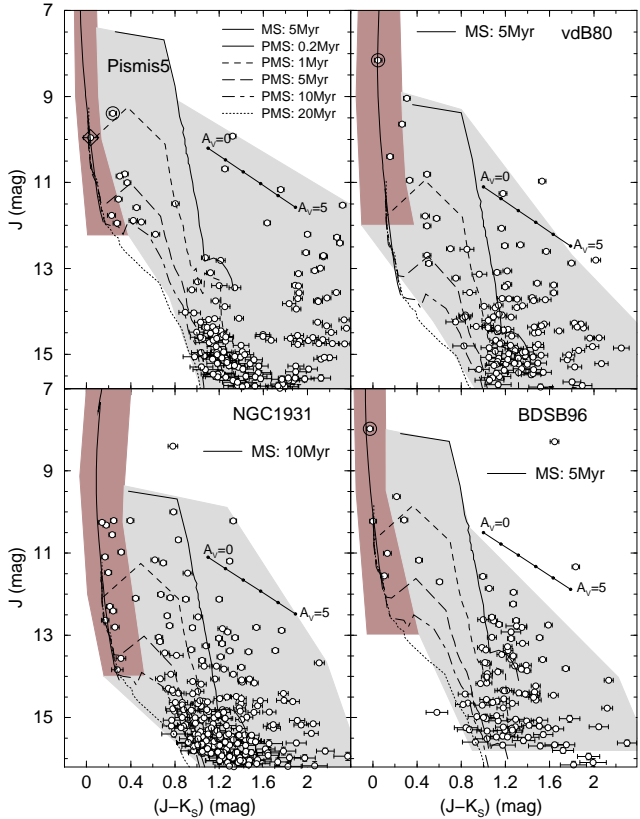}}
\caption[]{Adopted MS$+$PMS isochrone solutions to the decontaminated CMDs. SIMBAD bright 
stars and reddening vectors as in Figs.~\ref{fig4} and \ref{fig5}. Shaded polygons show 
the MS (dark-gray) and PMS (light-gray) colour-magnitude filters (Sect.~\ref{struc}).}
\label{fig6}
\end{figure}

Colour distributions wider than the spread predicted by PMS models occur in all cases 
(Fig.~\ref{fig6}), which reflects differential reddening. To examine this issue we 
show in Figs.~\ref{fig4}-\ref{fig6} reddening vectors computed with the 2MASS ratios 
(Sect.~\ref{2mass}) for visual absorptions $\aV=0~{\rm to}~5$. Taking into account the 
PMS isochrone fit (Fig.~\ref{fig6}), the differential reddening appears to be lower than 
$\aV=5$.

\begin{figure}
\resizebox{\hsize}{!}{\includegraphics{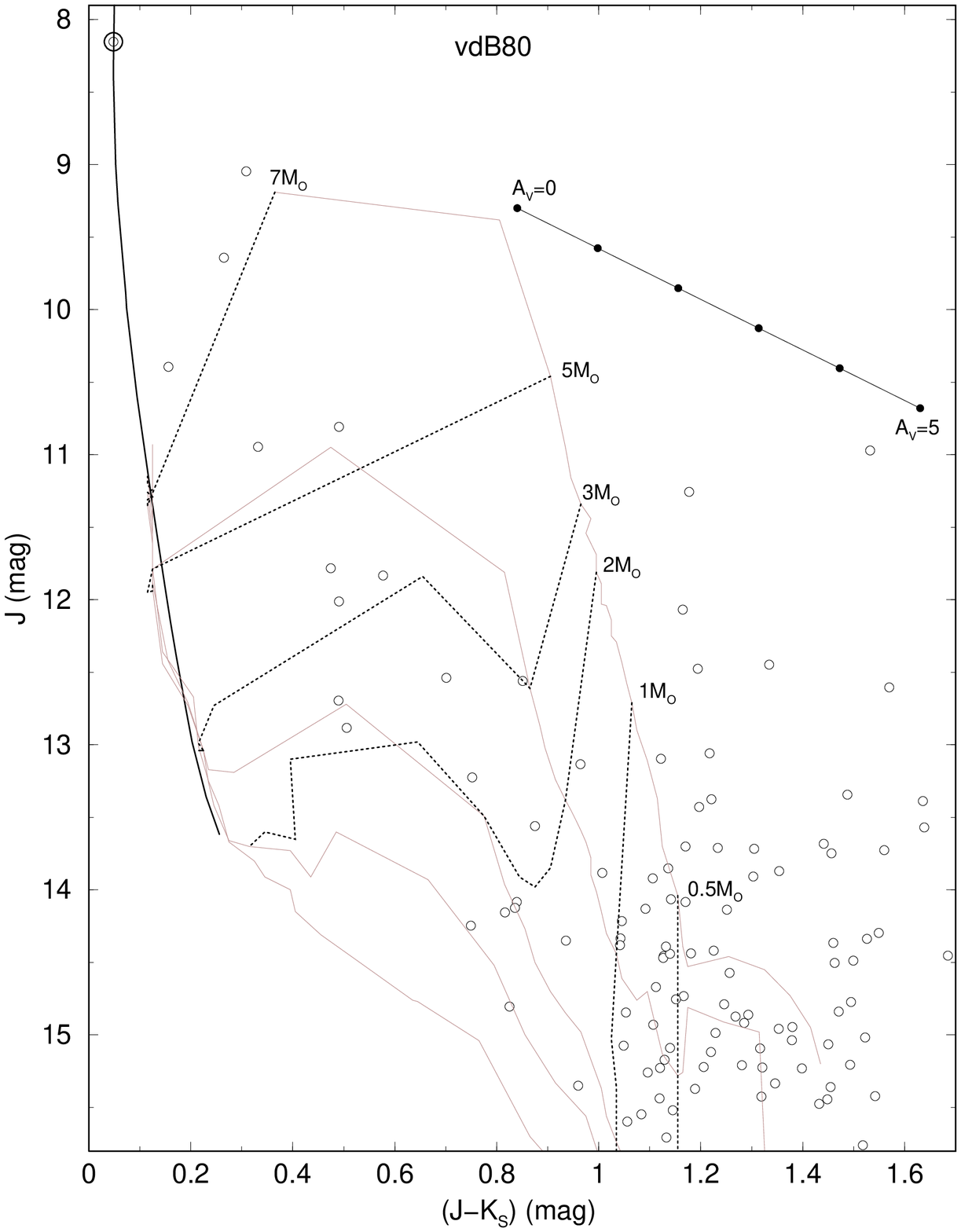}}
\caption[]{Decontaminated CMD of vdB\,80, used to illustrate the PMS mass estimate.
Light lines: MS and PMS isochrones set as in Fig.~\ref{fig6}. Heavy-dotted: PMS
evolutionary tracks for different masses.}
\label{fig7}
\end{figure}

We base the fundamental parameter derivation on the field-decontaminated CMD morphologies 
(Fig.~\ref{fig6}), using as constraint the combined MS and PMS star distribution. We 
adopt solar metallicity isochrones because the clusters are young and located not far from 
the Solar circle (see below), a region essentially occupied by $[Fe/H]\approx0.0$ OCs 
(\citealt{Friel95}).

To represent the MS we use Padova isochrones (\citealt{Girardi2002}) computed with the 2MASS 
\jj, \hh, and \ks\ filters\footnote{{\em http://stev.oapd.inaf.it/cgi-bin/cmd}. These 
isochrones are very similar to the Johnson-Kron-Cousins ones (e.g. \citealt{BesBret88}), with 
differences of at most 0.01 in colour (\citealt{TheoretIsoc}).}. The similar decontaminated 
CMD morphologies, typical of young ages (Fig.~\ref{fig6}), indicate similar age-solutions for
the present objects.

Sophisticated CMD fit approaches are available, especially for the MS (as summarised 
in \citealt{NJ06}). However, given the poorly-populated MSs, the 2MASS photometric 
uncertainties for the lower sequences and the important population of PMS stars, we 
decided for the direct comparison of isochrones with 
the decontaminated CMD morphology. Thus, fits are made {\em by eye}, taking the combined 
MS and PMS stellar distributions as constraint, allowing as well for differential reddening
and photometric uncertainties. Isochrones of \citet{Siess2000} with ages in the range 
0.2---20\,Myr are used to characterise the PMS sequences. The results are shown in 
Fig.~\ref{fig6} and discussed below. 

{\tt Pismis\,5:}
Given the poorly-populated MS, acceptable fits to the decontaminated MS morphology are 
obtained with any isochrone with age in the range 1---10\,Myr. The PMS stars are basically 
contained within the 0.2\,Myr and 10\,Myr isochrones as well, which implies a similar age 
range as the MS. Thus, we take the 5\,Myr isochrone as representative solution, with a 4\,Myr 
spread. In this case, $CD-39~4599$ is in the MS while $CD-39~4582$ would be a massive 
($\sim5\,\ms$) PMS almost reaching the MS.

\begin{figure}
\resizebox{\hsize}{!}{\includegraphics{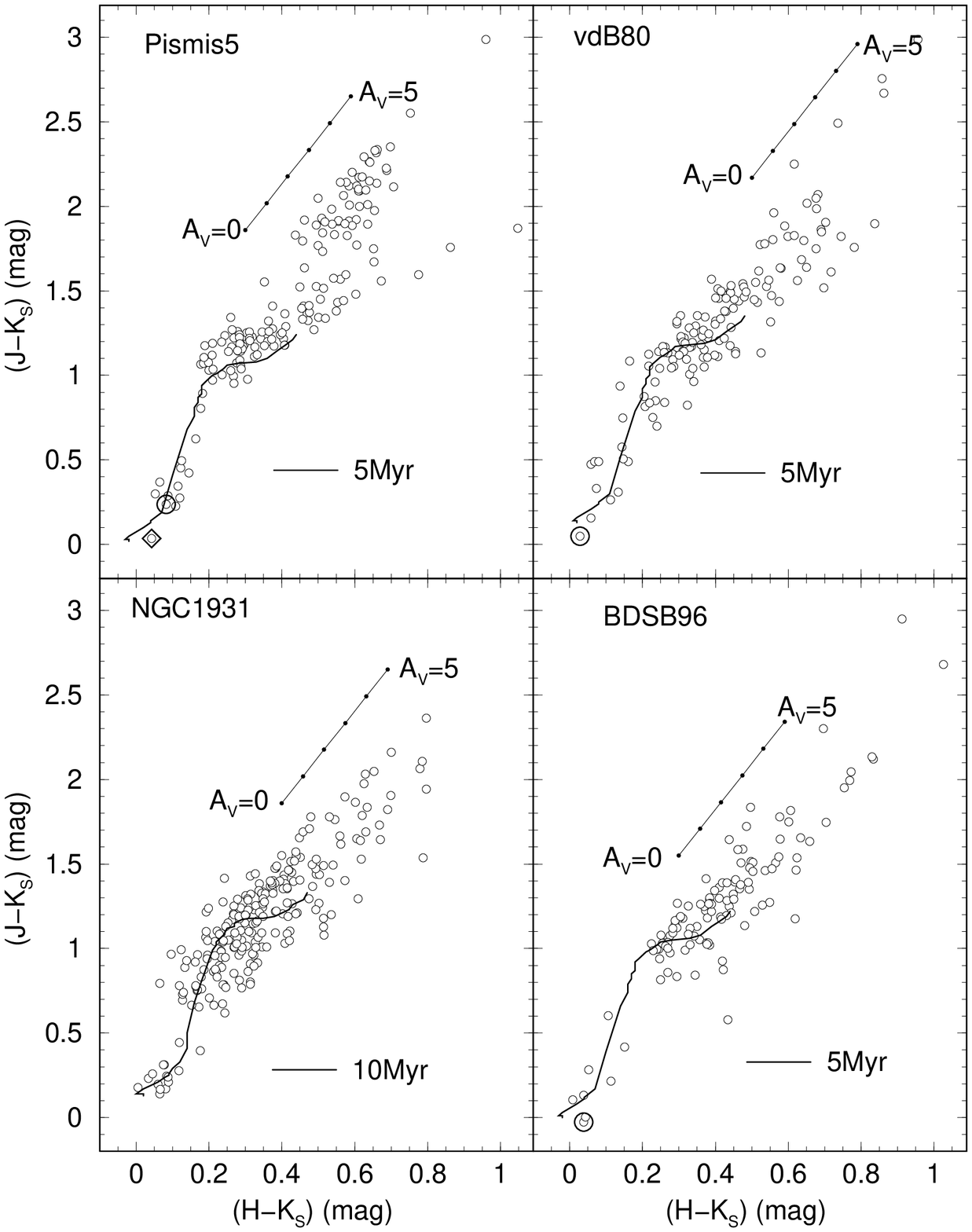}}
\caption[]{Decontaminated colour-colour diagrams. PMS isochrones set with the derived 
reddening values (Sect.~\ref{DFP}). SIMBAD bright stars and reddening vectors as in 
Figs.~\ref{fig4} and \ref{fig5}.}
\label{fig8}
\end{figure}

With the adopted solution, the fundamental parameters of Pismis\,5 are the near-IR reddening 
$\ejh=0.13\pm0.01$ ($\ebv=0.42\pm0.03$ or $A_V=1.3\pm0.1$), the observed and absolute distance 
moduli $\mMJ=10.4\pm0.1$ and $\mMo=10.04\pm0.10$, respectively, and the distance from the Sun 
$\ds=1.0\pm0.1$\,kpc. We adopt $\rs=7.2\pm0.3$\,kpc (\citealt{GCProp}) as the Sun's distance to 
the Galactic centre to compute Galactocentric distances\footnote{Derived by means of the Globular
Cluster (GC) spatial distribution. Recently, \citet{Trippe08} found $\dgc=8.07\pm0.32$\,kpc while 
\citet{Ghez08} found $\dgc=8.0\pm0.6$\,kpc or $\dgc=8.4\pm0.4$\,kpc, under different assumptions.}.
For $\rs=7.2$\,kpc, the Galactocentric distance of Pismis\,5 is $\dgc=7.5\pm0.1$\,kpc, which puts 
it $\approx0.3$\,kpc outside the Solar circle. This solution is shown in Fig.~\ref{fig6}. 

{\tt vdB\,80:} 
The bright star $NSV~2998$ (\jj=8.153) helps constrain the MS age to 
$5\pm2$\,Myr, which agrees with the $4.5\pm1.5$\,Myr age of \citet{Ahumada01}. 
Similarly to Pismis\,5, the PMS stars are essentially found within the 0.2---10\,Myr 
isochrones. With this solution, the fundamental parameters of vdB\,80 are 
$\ejh=0.19\pm0.03$ ($\ebv=0.61\pm0.10$ or $A_V=2.0\pm0.3$), $\mMJ=12.1\pm0.3$, 
$\mMo=11.58\pm0.31$, $\ds=2.1\pm0.3$\,kpc, and $\dgc=8.9\pm0.3$\,kpc, thus 
$\approx1.7$\,kpc outside the Solar circle.

{\tt NGC\,1931:} 
Besides the most populated MS of the present sample, NGC\,1931 presents the smallest 
gap between the faintest MS star and the PMS distribution, which suggest a more evolved 
phase. Indeed, the PMS stars distribute between the 0.2---20\,Myr isochrones. Based on 
this, the MS age can be set at $10\pm3$\,Myr, which agrees with the estimates of 
\citet{PM86}, \citet{Bhatt94}, \citet{Lata02} and \citet{CCS04}. This solution implies 
$\ejh=0.19\pm0.03$ ($\ebv=0.61\pm0.10$ or $A_V=2.0\pm0.3$), $\mMJ=12.4\pm0.3$, 
$\mMo=11.86\pm0.22$, $\ds=2.4\pm0.2$\,kpc, and $\dgc=9.6\pm0.2$\,kpc, thus $\approx2.4$\,kpc 
beyond the Solar circle. Our value for \ds\ is intermediate between the 1.8\,kpc of 
\citet{MFJ79} and 3.1\,kpc of \citet{CCS04}, and agrees with the 2.2\,kpc of \citet{PM86} 
and \citet{Bhatt94}.

{\tt BDSB\,96:}
The CMD of this object is similar to that of vdB\,80 - including the presence of a bright
star ($HD~53623$) in the MS, which suggests a similar solution, with the MS age $5\pm3$\,Myr. 
This solution implies $\ejh=0.12\pm0.03$ ($\ebv=0.37\pm0.10$ or $A_V=1.1\pm0.3$), 
$\mMJ=11.0\pm0.2$, $\mMo=10.68\pm0.22$, $\ds=1.4\pm0.2$\,kpc, and $\dgc=8.2\pm0.2$\,kpc, 
thus $\approx1.0$\,kpc outside the Solar circle. Our value of \ds\ agrees with the 1.1\,kpc
of \citet{BDS2003}.

Besides the MS age, the present objects have in common a significant age spread implied 
by the PMS stars. This indicates a non-instantaneous star formation process, similar to 
what was found in our previous studies of, e.g. NGC\,4755, NGC\,6611 and NGC\,2244.

\subsection{Colour-colour diagrams}
\label{2CD}

When transposed to the near-IR colour-colour diagram $\jk\times\hk$ (Fig.~\ref{fig8}), the 
age and reddening solutions derived for the present objects consistently match their field-star 
decontaminated photometry. Since they include PMS stars, we use tracks of \citet{Siess2000} to 
characterise the age. MS stars lie on the blue side of the diagram. As expected from the CMDs,
a significant fraction of the stars appears to be very reddened. Besides, since most stars
have \hk\ colours close to the isochrone, within the uncertainties, the fraction of stars still 
bearing circunstellar discs (with excess in \hh) appears to be low (\citealt{N4755} and
references therein). 

\section{Cluster structure}
\label{struc}

We use the RDPs, defined as the projected stellar number density around the cluster 
centre (i.e. the maximum of the surface density maps - Sect.~\ref{DecOut}), to derive 
structural parameters. To minimise noise, we work with colour-magnitude 
filters (shown in Fig.~\ref{fig6}) to exclude stars with colours unlike those of the 
cluster CMD morphology\footnote{They are wide enough to include cluster MS and PMS stars, 
together with the photometric uncertainties and binaries (and other multiple systems).}. 
This enhances the RDP contrast relative to the background, especially in crowded fields 
(e.g. \citealt{BB07}). Examples of the advantage of using colour-magnitude filters can be 
found in, e.g. \citet{OldOCs}, \citet{PlaNeb} and \citet{AntiC}. 

Rings of increasing width with distance from the cluster centre are used to preserve
spatial resolution near the centre and minimise noise at large radii. The set of ring 
widths used is $\Delta\,R=0.25,\ 0.5,\ 1.0,\ 2.0,\ {\rm and}\ 5\arcmin$, respectively 
for $0\arcmin\le R<0.5\arcmin$, $0.5\arcmin\le R<2\arcmin$, $2\arcmin\le R<5\arcmin$, 
$5\arcmin\le R<20\arcmin$, and $R\ge20\arcmin$. Because of the low number of stars in the
central parts of vdB\,80, we used $\Delta\,R=0.5$ for $R\le1\arcmin$. The residual background 
level of each RDP corresponds to the average number-density of field stars. The $R$ coordinate 
(and uncertainty) of each ring corresponds to the average position and standard deviation 
of the stars inside the ring. As a caveat we note that the present OCs are not spherical,
especially at the outskirts (Fig.~\ref{fig3}), which might affect the RDPs for large radii. 
Because of this, deviations in the central parts of the RDP are not expected to be 
significant. 

\begin{table*}
\caption[]{Derived structural parameters}
\label{tab2}
\renewcommand{\tabcolsep}{1.6mm}
\renewcommand{\arraystretch}{1.25}
\begin{tabular}{cccccccccccc}
\hline\hline
Cluster&$\sigma_{bg}$&$\sigma_0$&\rc&\rl&$\delta_c$&$1\arcmin$&$\sigma_{bg}$&$\sigma_0$&\rc&\rl\\
       &$\rm(*\,\arcmin^{-2})$&$\rm(*\,\arcmin^{-2})$&(\arcmin)&(\arcmin)& &(pc)&
$\rm(*\,pc^{-2})$&$\rm(*\,pc^{-2})$&(pc)&(pc)\\
(1)&(2)&(3)&(4)&(5)&(6)&(7)&(8)&(9)&(10)&(11)\\
\hline
Pismis\,5&$1.81\pm0.02$&$10.2\pm4.3$&$0.69\pm0.22$&$6.0\pm0.3$&$6.6\pm2.3$&0.296&$20.7\pm0.3$&$116.7\pm49.2$&
   $0.20\pm0.06$&$1.8\pm0.1$\\
   
vdB\,80&$1.25\pm0.02$&$20.3\pm5.0$&$0.46\pm0.08$&$5.8\pm0.3$&$17.2\pm4.0$&0.599&$3.5\pm0.1$&$56.4\pm14.0$&
   $0.28\pm0.05$&$3.5\pm0.2$\\
   
NGC\,1931&$2.08\pm0.02$&$19.6\pm6.1$&$0.70\pm0.15$&$8.5\pm0.5$&$10.4\pm2.9$&0.684&$4.4\pm0.1$&$41.9\pm13.0$&
   $0.48\pm0.10$&$5.8\pm0.3$\\
   
BDSB\,96&$1.29\pm0.01$&$28.2\pm11.8$&$0.88\pm0.30$&$11.0\pm1.0$&$22.9\pm9.2$&0.397&$8.1\pm0.1$&$178.8\pm74.8$&
   $0.35\pm0.12$&$4.4\pm0.4$\\
\hline
\end{tabular}
\begin{list}{Table Notes.}
\item Core (\rc) and cluster (\rl) radii are given in angular and absolute units. 
Col.~6: cluster/background density contrast parameter ($\delta_c=1+\sigma_0/\sigma_{bg}$). 
Col.~7: arcmin to parsec scale. 
\end{list}
\end{table*}

\begin{figure}
\resizebox{\hsize}{!}{\includegraphics{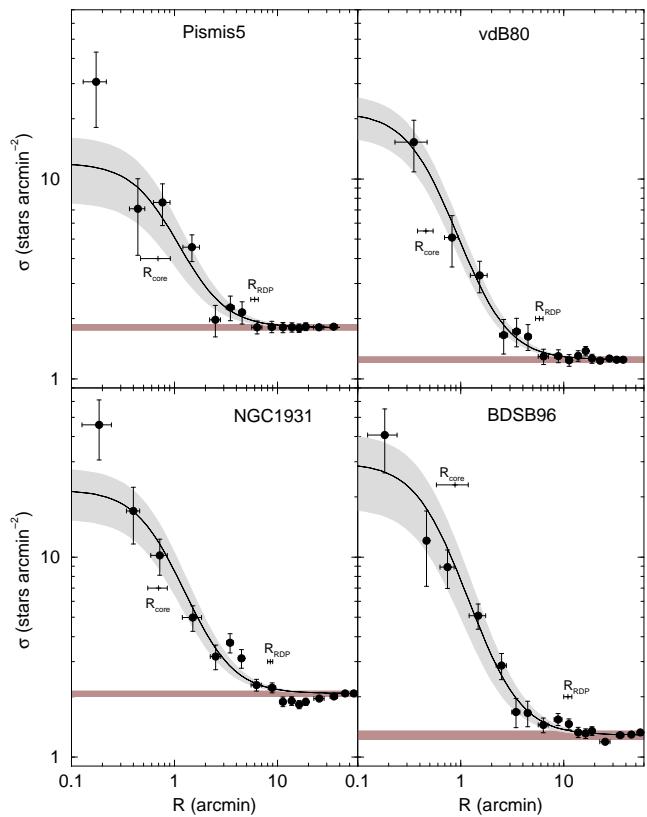}}
\caption[]{Stellar RDPs built with colour-magnitude filtered photometry together with
the best-fit King-like profile (solid line), the $1\sigma$ uncertainty (light-shaded
region) and the background level (shaded polygon). Note the central cusp in the RDPs 
of Pismis\,5 and NGC\,1931.}
\label{fig9}
\end{figure}

Minimisation of the number of non-cluster stars by the colour-magnitude filters yielded 
RDPs (Fig.~\ref{fig9}) highly contrasted relative to the background. For simplicity we 
fit the RDPs with $\sigma(R)=\sigma_{bg}+\sigma_0/(1+(R/R_c)^2)$, where $\sigma_{bg}$ is 
the residual background density, $\sigma_0$ is the central density of stars, and \rc\ is the 
core radius. This function, applied to star counts, is similar to that introduced by 
\cite{King1962} to describe the surface-brightness profiles in the central parts of 
GCs\footnote{Besides, because of the relatively small number of stars in the present OCs, 
fluctuations in surface-brightness profiles are expected to be larger than those in RDPs. 
Alternative RDP fit functions are discussed in \citet{StrucPar}.}. To minimise degrees of 
freedom, $\sigma_0$ and \rc\ follow from the fit, while $\sigma_{bg}$ is measured in the field 
and kept fixed. The best-fit solutions and uncertainties are shown in Fig.~\ref{fig9}, and the 
parameters in Table~\ref{tab2}. We also estimate the cluster radius (\rl), i.e. the 
distance from the centre where the cluster RDP and field fluctuations are statistically 
indistinguishable (e.g. \citealt{DetAnalOCs}), and the density contrast parameter 
$\delta_c=1+\sigma_0/\sigma_{bg}$ (Table~\ref{tab2}). Interestingly, the density-contrast 
parameter reaches high values, $7\la\delta_c\la23$, as expected from compact star clusters.

Within uncertainties, the adopted King-like function describes the RDPs of vdB\,80 and BDSB\,96 
along the whole radius range. To a lesser degree, the same applies for Pismis\,5 and NGC\,1931, 
in which the innermost bin presents an excess over the fit. In old star clusters, such a cusp 
has been attributed to a post-core collapse structure, like those detected in some GCs (e.g. 
\citealt{TKD95}). Gyr-old OCs, e.g. NGC\,3960 (\citealt{N3960}) and LK\,10 (\citealt{LKstuff}), 
also display such feature, which is related to dynamical evolution. With respect to very young 
clusters, the RDPs of NGC\,2244 (\citealt{N2244}) and NGC\,6823 (\citealt{Bochum1}) also present 
a central cusp, similarly to those in Pismis\,5 and NGC\,1931. Clusters are not expected to 
dynamically evolve into a post-core collapse on such short time-scales. Consequently, the cusp 
in young clusters must be related to molecular cloud fragmentation and/or star-forming effects, 
and may suggest early deviation from dynamical equilibrium (Sect.~\ref{Discus}).

Taken at face value, the core radii of the present objects ($\rm 0.2\la\rc(pc)\la0.5$) 
fall on the low-\rc\ tail of the distribution derived for a sample of relatively 
nearby OCs by \citet{Piskunov07}.

\section{Mass estimate}
\label{MF}

The most conspicuous CMD feature of the present OCs is the dominant presence
of PMS stars, followed by the poorly-populated MS (Fig.~\ref{fig6}). Obviously,
in each cluster, most of the mass is stored in the PMS stars. Because 
the decontamination algorithm actually excludes stars (in integer numbers), the 
decontaminated photometry should be used essentially to determine the CMD morphology 
or the colour-colour distribution (Sect.~\ref{DFP}). However, when magnitude (or 
alternatively mass) bins are used to build the MFs, the bin-to-bin subtraction of 
the comparison field (normalised to the same projected area as the cluster) 
contribution is expected to produce fractional numbers, which should be incorporated 
into the MFs (e.g. \citealt{DetAnalOCs}). For this reason we use the colour-magnitude 
filtered photometry (Fig.~\ref{fig6}) to estimate the cluster mass.

The number of MS stars is derived by counting the stars within $R\le\rl$ (in bins of 
$\Delta\jj=1$\,mag), and subtracting those in the field (normalised to the same area). 
The corresponding stellar mass in each magnitude bin is taken from the mass-luminosity 
relation derived from the isochrone fit (Sect.~\ref{DFP}). Summing the values in each 
magnitude bin produces the total number ($n_{MS}$) and mass ($m_{MS}$) of MS stars. A 
similar strategy is applied to the PMS stars. We consider the evolutionary tracks for 
the PMS masses 0.5, 1, 2, 3, 5 and 7\,\ms\ (Fig.~\ref{fig7}). The number of PMS stars 
between any 2 tracks is counted for $R\le\rl$ and the field, taking into account the 
reddening vectors. The average mass between two evolutionary tracks is taken as the 
mass of a respective PMS star. All tracks summed result in the number ($n_{PMS}$) and 
mass ($m_{PMS}$) of PMS stars (Table~\ref{tab3}).

\begin{table*}
\caption[]{MS and PMS stellar content}
\label{tab3}
\renewcommand{\tabcolsep}{3.0mm}
\renewcommand{\arraystretch}{1.25}
\begin{tabular}{ccccccccccc}
\hline\hline
&\multicolumn{3}{c}{MS}&&\multicolumn{2}{c}{PMS}&&\multicolumn{3}{c}{MS$+$PMS}\\
\cline{2-4}\cline{6-7}\cline{9-11}
Cluster&$\Delta\,m$&$n_{MS}$&$m_{MS}$&&$n_{PMS}$&$m_{PMS}$&&$n_{MS+PMS}$&$m_{MS+PMS}$&$\chi$\\
&(\ms)&(stars)&(\ms)&&(stars)&(\ms)&&(stars)&(\ms)&\\
\hline
(1)&(2)&(3)&(4)&&(5)&(6)&&(7)&(8)&(9)\\
\hline
Pismis\,5&1.9-13.4&$3\pm1$&$12\pm3$&&$100\pm20$&$46\pm7$&&$103\pm21$&$58\pm8$&$-0.03\pm0.19$\\
   
vdB\,80&2.3-21.6&$4\pm1$&$40\pm11$&&$108\pm12$&$55\pm13$&&$112\pm13$&$95\pm17$&$-0.24\pm0.09$\\
   
NGC\,1931&2.0-17.8&$15\pm3$&$62\pm13$&&$183\pm33$&$115\pm37$&&$198\pm34$&$177\pm39$&$-0.02\pm0.08$\\
   
BDSB\,96&1.8-23.8&$5\pm1$&$23\pm6$&&$171\pm23$&$131\pm21$&&$176\pm24$&$154\pm23$&$+0.80\pm0.08$\\
\hline
\end{tabular}
\begin{list}{Table Notes.}
\item Col.~2: Effective MS mass range. Col.~9: MS$+$PMS mass function slope $\chi$,
derived from the fit of $\phi(m)\propto m^{-(1+\chi)}$.
\end{list}
\end{table*}

As anticipated by the CMDs (Fig.~\ref{fig6}), the MSs as a rule, are poorly-populated. Pismis\,5, 
vdB\,80 and BDSB\,96 contain about 4 MS stars, while NGC\,1931 hosts about 15 MS stars. Such low 
numbers reflect the young age and low-mass nature of these objects. Indeed, the mass of the 
present objects, as derived from their MS$+$PMS members, are within $60-180\,\ms$\footnote{As a 
caveat we note that, because of the differential reddening and the 2MASS photometric limit, we
may not detect the PMS stars of very-low mass; thus the mass values may be somewhat higher.}. 
For comparison, Bochum\,1, NGC\,6823, NGC\,2244 and NGC\,2239 contain 128, 65, 26 and 26 MS 
stars, respectively, while their (MS$+$PMS) masses are 720, 1150, 625 and 301\,\ms, 
respectively. 

\begin{figure}
\resizebox{\hsize}{!}{\includegraphics{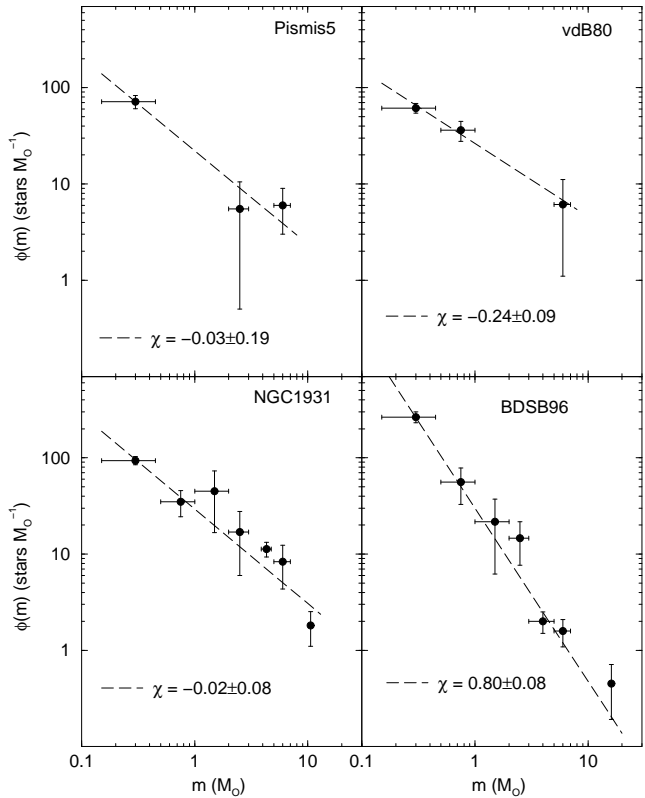}}
\caption[]{MS$+$PMS mass functions fitted by the function $\phi(m)\propto m^{-(1+\chi)}$ 
(dashed line).}
\label{fig10}
\end{figure}

An estimate of the MS and PMS MFs can be made with the approach described above. In 
Fig.~\ref{fig10} we show the resulting MFs for the PMS and MS stars combined. Since the 
MSs are very-poorly populated (Table~\ref{tab3}), it should be noted that these MFs 
correspond essentially to the PMS stars; the MSs contribute somewhat to the range 
$m\ga2\,\ms$. Besides, given the simplifying assumptions, the error bars in the MFs
are only formal, and should be taken as a lower limit. Bearing this in mind, the MFs 
can be represented by the function $\phi(m)=\frac{dN}{dm}\propto m^{-(1+\chi)}$, with slopes 
$\chi$ (Table~\ref{tab3}) significantly flatter than the $\chi=1.35$ of \citet{Salpeter55} 
initial mass function (IMF). Interestingly, these MFs do not appear to present a turnover 
at the sub-Solar mass range, as suggested by the IMFs of \citet{Kroupa2001} and \citet{Chab03}. 
However, given the low number of stars and the simplifying
assumptions, we have omitted several possible systematic biases associated with the IMF 
determination: {\em (i)} uncertainties in the low-mass IMF due to differential reddening 
and possibly crowding (the latter should be minimal, since the objects are intrinsically
poorly populated), {\em (ii)} missing companion stars (e.g. \citealt{MA08}), {\em (iii)} 
binning (e.g. \citealt{MA05}), and {\em (iv)} errors on the mass/luminosity relation of 
individual stars.  

\section{Discussion} 
\label{Discus}

To put Pismis\,5, vdB\,80, NGC\,1931 and BDSB\,96 into perspective, their astrophysical 
parameters (Sects.~\ref{DFP}, \ref{struc}, \ref{MF}) are compared to those of a sample of 
nearby OCs in different environments. The reference sample contains some bright nearby OCs 
(\citealt{DetAnalOCs}; \citealt{N4755}) together with a group of OCs projected towards the 
central parts of the Galaxy (\citealt{BB07}). The young OCs NGC\,6611 (\citealt{N6611}), 
NGC\,6823 (\citealt{Bochum1}) and NGC\,2239 (\citealt{N2244}) are included for comparison 
with gravitationally bound objects of similar ages, while Bochum\,1 (\citealt{Bochum1}) and 
NGC\,2244 (\citealt{N2244}) might be dynamically evolving into OB associations. All objects
have been similarly analysed.

\begin{figure}
\resizebox{\hsize}{!}{\includegraphics{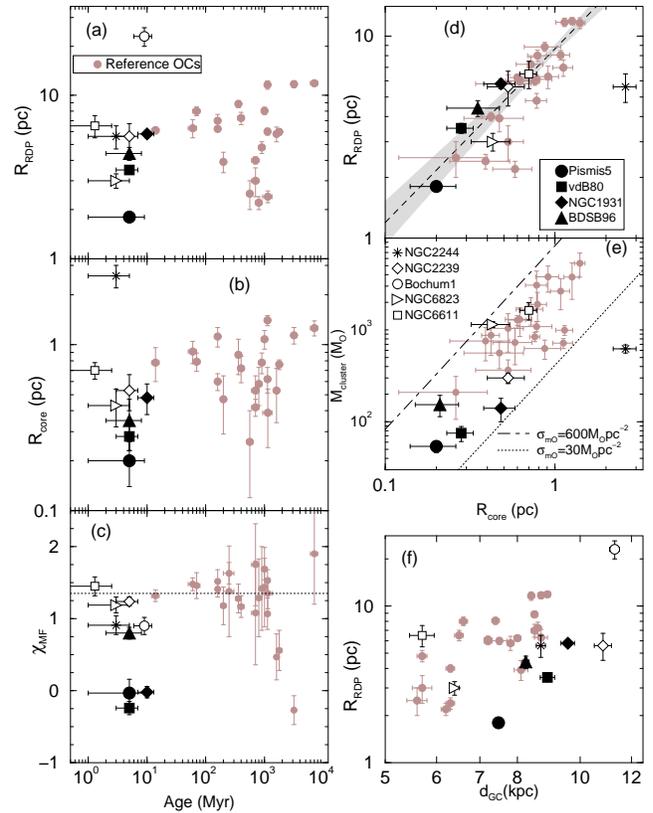}}
\caption{Diagrams dealing with astrophysical parameters of OCs. Gray-shaded circles:
reference OCs. The young star clusters NGC\,6611, NGC\,6823, Bochum\,1, NGC\.2244 and
NGC\,2239 are indicated for comparison purposes. Analytical relations in panels (d) and 
(e) are discussed in the text. Dotted line in panel (c): Salpeter's IMF slope $\chi=1.35$.}
\label{fig11}
\end{figure}

We work with the diagnostic-diagrams shown in Fig.~\ref{fig11}. Panels (a) and (b) examine 
the dependence of cluster (\rl) and core (\rc) radii on age, respectively. Pismis\,5, vdB\,80
and BDSB\,96 (and to a lesser degree NGC\,1931) present small \rc\ and \rl, similar to NGC\,6823 
and typical of Gyr-old OCs 
undergoing disruption near the Solar circle (e.g. \citealt{Bergond2001}; \citealt{Lamers05}).
Within uncertainties, the 4 objects follow the relation $\rl=(8.9\pm0.3)\times
R_{\rm C}^{(1.0\pm0.1)}$ (panel d), derived with the reference sample. They also appear
to follow the relation of increasing OC size with Galactocentric distance\footnote{It may be 
partly primordial, in the sense that the high molecular gas density in central Galactic regions 
may have produced small clusters (e.g. \citealt{vdB91}).} (panel f).

When the mass-density radial distribution follows a King-like profile (e.g. \citealt{OldOCs};
\citealt{StrucPar}; \citealt{PlaNeb}), the mass inside \rl\ is a function of \rc\ and the 
central mass-surface density ($\sigma_{M0}$), $\rm M_{clus}=\pi\,R^2_{C}\sigma_{M0}\ln\left
[1+\left(\rl/\rc\right)^2\right]$. With the relation between \rc\ and \rl\ (panel d), this 
equation becomes $\rm M_{clus}\approx13.8\sigma_{M0}\,R^2_{C}$. The reference OCs, together 
with the present sample, are contained within King-like distributions with $\sigma_{M0}$ 
constrained within $\rm30\la\sigma_{M0}\,(\ms\,pc^{-2})\la600$ (panel e). The exception is 
NGC\,2244, which appears to present too big a core for the total mass. 

Finally, when the total (MS$+$PMS) MF slope is considered (panel c), Pismis\,5, vdB\,80 and 
NGC\,1931 present MFs significantly flatter than those of similarly young OCs. To a lesser
degree, the same applies to the newly formed BDSB\,96. Such flat slopes are equivalent to those 
derived for some very old OCs in the reference sample that, in general, undergo advanced 
dynamical evolution (e.g. \citealt{DetAnalOCs}). Given the young age, low mass, irregular 
RDP and the flat mass function, the OCs NGC\,1931 and Pismis\,5 may be evolving into 
OB associations or remnants in a few $10^7$\,yr. Since both OCs are young, such effects may
be related to another mechanism than age-related dynamical evolution. It is probably associated 
with star formation. vdB\,80 and BDSB\,96, on the other hand, appear to be typical young and 
low-mass OCs.

\section{Summary and conclusions}
\label{Conclu}

Probably because of the interplay between environmental conditions, star-formation 
efficiency and stellar mass, only a few percent of the embedded clusters survive
the first few tens of Myr. Thus, the derivation of astrophysical parameters of OCs 
in this phase may shed some light on the roles played by the above process in the 
dissolution/survival issue. In this context, poorly-populated and massive OCs in 
different environments are yet to be investigated.

In the present paper we derive astrophysical parameters and investigate the nature of 
the young and low-mass OCs Pismis\,5, vdB\,80, NGC\,1931 and BDSB\,96.
The wide-field and near-IR depth provided by 2MASS (with errors lower than 0.1\,mag) 
are employed coupled to field-star decontaminated photometry. This enhances cluster 
CMD evolutionary sequences and stellar radial density profiles, yielding more 
constrained fundamental and structural parameters. The OCs are located within 
$174\degr\la\ell\la260\degr$ and $|b|\la9\degr$, thus, the errors potentially induced 
by the background star subtraction are not critical.

The decontaminated CMDs exhibit similar properties, basically a poorly-populated
MS, a dominant fraction of PMS stars together with some differential reddening.
MS ages are constrained within $5\pm4$\,Myr (Pismis\,5, vdB\,80, BDSB\,96) 
and $10\pm3$\,Myr (NGC\,1931). However, the PMS stars suggest a wider age spread
($\sim20$\,Myr), consistent with a non-instantaneous star formation process. The 
total (MS$+$PMS) stellar masses are low, within $\sim60-180\,\ms$, with mass functions
significantly flatter than Salpeter's IMF.

The present OCs are rather small ($\rc\la0.48$\,pc, $\rl\la5.8$\,pc), 
particularly Pismis\,5 with $\rc\approx0.2$\,pc and $\rl\approx1.8$\,pc. Structurally, 
the (MS$+$PMS) stellar RDPs follow a cluster-like profile for most of the radius range. 
Exceptions are the inner regions of NGC\,1931 and especially Pismis\,5, which present 
a marked stellar-density excess. At $\sim10$\,Myr, such a central cusp cannot result 
from large-scale cluster dynamical evolution. Instead, it probably is associated with 
molecular cloud fragmentation and/or star-formation effects.

BDSB\,96 and vdB\,80 present structural properties of typical young, low-mass 
OCs, although with flat mass functions. On the other hand, the irregular 
RDPs - together with the low cluster mass and flat mass functions - of NGC\,1931 
and especially Pismis\,5, suggest that both OCs deviate from dynamical equilibrium.
They are possibly evolving to become OB associations or remnants in a few $10^7$\,yr.

\section*{Acknowledgements}
We thank the anonymous referee for interesting suggestions.
We acknowledge support from the Brazilian Institution CNPq.
This publication makes use of data products from the Two Micron All Sky Survey, which
is a joint project of the University of Massachusetts and the Infrared Processing and
Analysis Centre/California Institute of Technology, funded by the National Aeronautics
and Space Administration and the National Science Foundation. This research has made 
use of the WEBDA database, operated at the Institute for Astronomy of the University
of Vienna.

\label{lastpage}

\begin{thebibliography}{}

\bibitem[\protect\citeauthoryear{Ahumada et al.}{2001}]{Ahumada01}
   Ahumada A.V., Clari\'a J.J., Bica E., Dutra C.M. \& Torres M.C.
   2001, A\&A, 377, 845
   
\bibitem[\protect\citeauthoryear{van den Bergh}{1966}]{vdB66}
   van den Bergh S. 1966, AJ, 71, 990

\bibitem[\protect\citeauthoryear{van den Bergh, Morbey \& Pazder}{1991}]{vdB91}
   van den Bergh S., Morbey C. \& Pazder J. 1991, ApJ, 375, 594
   
\bibitem[\protect\citeauthoryear{Bergond, Leon \& Guilbert}{2001}]{Bergond2001}
   Bergond G., Leon S. \& Guilbert J. 2001, A\&A, 377, 462
     
\bibitem[\protect\citeauthoryear{Bessel \& Brett}{1988}]{BesBret88}
   Bessel M.S. \& Brett J.M. 1988, PASP, 100, 1134
   
\bibitem[\protect\citeauthoryear{Bhatt et al.}{1994}]{Bhatt94}
   Bhatt B.C., Pandey A.K., Mahra H.S. \& Paliwal D.C. 1994, BASI, 22, 291
   
\bibitem[\protect\citeauthoryear{Bica et al.}{2003}]{BDS2003}
   Bica E., Dutra C.M., Soares, J. \& Barbuy B. 2003, A\&A, 404, 223
   
\bibitem[\protect\citeauthoryear{Bica, Bonatto \& Dutra}{2008}]{Bochum1}
   Bica E., Bonatto C. \& Dutra C. 2008, A\&A, 489, 1129

\bibitem[\protect\citeauthoryear{Bica et al.}{2006}]{GCProp}
   Bica E., Bonatto C., Barbuy B. \& Ortolani S. 2006, A\&A, 450, 105

\bibitem[\protect\citeauthoryear{Bica \& Bonatto}{2008}]{F1603}
   Bica E. \& Bonatto C. 2008, MNRAS, 384, 1733

\bibitem[\protect\citeauthoryear{Bica, Bonatto \& Camargo}{2008}]{ProbFSR}
   Bica E., Bonatto C. \& Camargo D. 2008, MNRAS, 385, 349

\bibitem[\protect\citeauthoryear{Bonatto \& Bica}{2003}]{M67}
   Bonatto C. \& Bica E. 2003, A\&A, 405, 525

\bibitem[\protect\citeauthoryear{Bonatto, Bica \& Girardi}{2004}]{TheoretIsoc}
   Bonatto C., Bica E. \& Girardi L. 2004, A\&A, 415, 571

\bibitem[\protect\citeauthoryear{Bonatto \& Bica}{2005}]{DetAnalOCs}
   Bonatto C. \&  Bica E. 2005, A\&A, 437, 483

\bibitem[\protect\citeauthoryear{Bonatto, Santos Jr. \& Bica}{2006}]{N6611}
   Bonatto C., Santos Jr. J.F.C. \& Bica E. 2006, A\&A, 445, 567

\bibitem[\protect\citeauthoryear{Bonatto et al.}{2006a}]{discProp}
   Bonatto C., Kerber L.O., Bica E. \& Santiago B.X. 2006a, A\&A, 446, 121

\bibitem[\protect\citeauthoryear{Bonatto et al.}{2006b}]{N4755}
   Bonatto C., Bica E., Ortolani S. \& Barbuy B. 2006b, A\&A, 453, 121
   
\bibitem[\protect\citeauthoryear{Bonatto \& Bica}{2006}]{N3960}
   Bonatto C. \& Bica E. 2006, A\&A, 455, 931

\bibitem[\protect\citeauthoryear{Bonatto \& Bica}{2007a}]{BB07}
   Bonatto C. \& Bica E. 2007a, MNRAS, 377, 1301

\bibitem[\protect\citeauthoryear{Bonatto \& Bica}{2007b}]{OldOCs}
   Bonatto C. \& Bica E. 2007b, A\&A, 473, 445
   
\bibitem[\protect\citeauthoryear{Bonatto \& Bica}{2008a}]{StrucPar}
   Bonatto C. \& Bica E. 2008a, A\&A, 477, 829
   
\bibitem[\protect\citeauthoryear{Bonatto, Bica \& Santos Jr.}{2008}]{PlaNeb}
   Bonatto C., Bica E. \& Santos Jr. J.F.C. 2008, MNRAS, 386, 324   
      
\bibitem[\protect\citeauthoryear{Bonatto \& Bica}{2008b}]{AntiC}
   Bonatto C. \& Bica E. 2008b, A\&A, 485, 81
   
\bibitem[\protect\citeauthoryear{Bonatto \& Bica}{2009a}]{LKstuff}
   Bonatto C. \& Bica E. 2009a, MNRAS, 392, 483
   
\bibitem[\protect\citeauthoryear{Bonatto \& Bica}{2009b}]{N2244}
   Bonatto C. \& Bica E. 2009b, MNRAS, accepted (astro-ph/0901.0833)

\bibitem[\protect\citeauthoryear{Cardelli, Clayton \& Mathis}{1989}]{Cardelli89}
   Cardelli J.A., Clayton G.C. \& Mathis, J.S. 1989, ApJ, 345, 245
   
\bibitem[\protect\citeauthoryear{Chabrier}{2003}]{Chab03}
   Chabrier G. 2003, PASP, 115, 763
   
\bibitem[\protect\citeauthoryear{Chen, Chen \& Shu}{2004}]{CCS04}
   Chen W.P., Chen C.W. \& Shu C.G. 2004, AJ, 128, 2306

\bibitem[\protect\citeauthoryear{Dutra, Santiago \& Bica}{2002}]{DSB2002}
   Dutra C.M., Santiago B.X. \& Bica E. 2002, A\&A, 383, 219
      
\bibitem[\protect\citeauthoryear{Friel}{1995}]{Friel95}
   Friel E.D. 1995, ARA\&A 1995, 33, 381
   
\bibitem[\protect\citeauthoryear{Ghez et al.}{2008}]{Ghez08}
   Ghez A.M., Salim S., Weinberg N.N., Lu J.R., Do T., Dunn J.K., Matthews K.,
   Morris M.R. et al. 2008, ApJ, 689, 1044
   
\bibitem[\protect\citeauthoryear{Girardi et al.}{2002}]{Girardi2002}
   Girardi L., Bertelli G., Bressan A., Chiosi C., Groenewegen M.A.T.,
   Marigo P., Salasnich B. \& Weiss A. 2002, A\&A, 391, 195  
     
\bibitem[\protect\citeauthoryear{Goodwin \& Bastian}{2006}]{GoBa06}
   Goodwin S.P. \& Bastian N. 2006, MNRAS, 373, 752
   
\bibitem[\protect\citeauthoryear{King}{1962}]{King1962}
   King I. 1962, AJ, 67, 471
   
\bibitem[\protect\citeauthoryear{Kroupa}{2001}]{Kroupa2001}
   Kroupa P. 2001, MNRAS, 322, 231
      
\bibitem[\protect\citeauthoryear{Lada \& Lada}{2003}]{LL2003}
   Lada C.J. \& Lada E.A. 2003, ARA\&A, 41, 57
   
\bibitem[\protect\citeauthoryear{Lamers et al.}{2005}]{Lamers05}
   Lamers H.J.G.L.M., Gieles M., Bastian N., Baumgardt H.,
   Kharchenko N.V. \& Portegies Zwart S. 2005, A\&A, 441, 117
   
\bibitem[\protect\citeauthoryear{Lata et al.}{2002}]{Lata02}
   Lata S., Pandey A.K., Sagar R. \& Mohan V. 2002, A\&A, 388, 158
   
\bibitem[\protect\citeauthoryear{Ma\'\i z Apell\'aniz}{2005}]{MA05}
   Ma\'\i z Apell\'aniz J. 2005, ApJ, 629, 873
   
\bibitem[\protect\citeauthoryear{Ma\'\i z Apell\'aniz}{2008}]{MA08}
   Ma\'\i z Apell\'aniz J. 2008, ApJ, 677, 1278
   
\bibitem[\protect\citeauthoryear{Massey, Johnson \& Gioia-Eastwood}{1995}]{Massey95}
   Massey P., Johnson K.E. \& De Gioia-Eastwood K. 1995, ApJ, 454, 151
   
\bibitem[\protect\citeauthoryear{Moffat, Fitzgerald \& Jackson}{1979}]{MFJ79}
   Moffat A.F.J., Fitzgerald M. \& Jackson P.D. 1979, A\&ASS, 38, 197
   
\bibitem[\protect\citeauthoryear{Naylor \& Jeffries}{2006}]{NJ06}
   Naylor T. \& Jeffries R.D. 2006, MNRAS, 373, 1251
   
\bibitem[\protect\citeauthoryear{Pandey \& Mahra}{1986}]{PM86}
   Pandey A.K. \& Mahra H.S. 1986, Ap\&SS, 120, 107
   
\bibitem[\protect\citeauthoryear{Piskunov et al.}{2007}]{Piskunov07}
   Piskunov A.E., Schilbach E., Kharchenko N.V., R\"oser S. \& Scholz R.-D.
   2007, A\&A, 468, 151
   
\bibitem[\protect\citeauthoryear{Pismis}{1959}]{Pismis59}
   Pismis P. 1959, BOTT, 2, 37
   
\bibitem[\protect\citeauthoryear{Salpeter}{1955}]{Salpeter55}
   Salpeter E. 1955, ApJ, 121, 161
   
\bibitem[\protect\citeauthoryear{Siess, Dufour \& Forestini}{2000}]{Siess2000}
   Siess L., Dufour E. \& Forestini M. 2000, A\&A, 358, 593

\bibitem[\protect\citeauthoryear{Skrutskie et al.}{1997}]{2mass1997}
   Skrutskie M., Schneider S.E., Stiening R., Strom S.E., Weinberg M.D.,
   Beichman C., Chester T., Cutri R .et al. 1997, in {\it The Impact
   of Large Scale Near-IR Sky Surveys}, ed. F. Garzon et al., Kluwer 
   (Netherlands), 210, 187
   
\bibitem[\protect\citeauthoryear{Trager, King \& Djorgovski}{1995}]{TKD95}
   Trager S.C., King I.R. \& Djorgovski S. 1995, AJ, 109, 218
   
\bibitem[\protect\citeauthoryear{Trippe et al.}{2008}]{Trippe08}
   Trippe S., Gillessen S., Gerhard O.E., Bartko H., Fritz T.K., Maness H.L.
   Eisenhauer F., Martins F., et al.  2008, A\&A, 492, 419
   
\bibitem[\protect\citeauthoryear{Vogt \& Moffat}{1973}]{VM73}
   Vogt N. \& Moffat A.F.J. 1973, A\&AS, 9, 97
   
\end{thebibliography}
\end{document}